\documentclass[aps,prd,showpacs,preprintnumbers,nofootinbib,floatfix,onecolumn]{revtex4}

\usepackage{bm}
\usepackage{latexsym}
\usepackage{dcolumn}
\usepackage{amsfonts,amssymb}
\usepackage{graphicx,epsfig}
\usepackage{psfrag}
\usepackage{subfigure}

\def\beq{\begin{equation}}
\def\eeq{\end{equation}}
\def\br{\begin{eqnarray}}
\def\er{\end{eqnarray}}
\def\pa{{\partial}}
\def\l{\left}
\def\r{\right}    
\def\eq#1{{Eq.~(\ref{#1})}}
\def\frab#1#2{\left(\frac{#1}{#2}\right)}

\begin{document}
  

\title{Particle creation, classicality and related issues in\\ quantum field theory: I. Formalism and toy models}
\author{Gaurang Mahajan}
\email[]{gaurang@iucaa.ernet.in}
\affiliation{IUCAA, Post Bag 4, Ganeshkhind, Pune - 411 007, India\\}
\author{T. Padmanabhan}
\email[]{nabhan@iucaa.ernet.in}
\affiliation{IUCAA, Post Bag 4, Ganeshkhind, Pune - 411 007, India\\}

\date{\today}


\begin{abstract} 
The quantum theory of a harmonic oscillator with a time dependent frequency arises in several important physical problems, especially in the study of quantum field theory in an external background. While the mathematics of this system is straightforward, several conceptual issues arise in such a study. We present a general formalism to address some of the conceptual issues like the emergence of classicality, definition of particle content, back reaction etc. In particular, we  parametrize the wave function in terms of a complex number (which we call excitation parameter) and express all physically relevant quantities in terms it. Many of the notions -- like those of particle number density, effective Lagrangian etc., which are usually defined using asymptotic in-out states -- are generalized as time-dependent concepts and we show that these generalized definitions
lead to useful and  reasonable results. Having developed the general formalism we apply it to several examples. Exact analytic expressions are found for a particular toy model and 
approximate analytic solutions are obtained in the extreme cases of adiabatic and highly non-adiabatic evolution. We then work out the exact results numerically for a variety of models and compare them with the analytic results and approximations.
  The formalism is  useful in addressing the question of emergence of classicality of the quantum state, its relation to particle production and to clarify several conceptual issues related to this. In Paper II (arXiv:0708.1237) which is a sequel to this, the formalism will be applied to analyze the corresponding issues in the context of quantum field theory in background cosmological models and electric fields.
\end{abstract}

\maketitle


\section{Introduction and Motivation}

The quantum harmonic oscillator with a time-dependent frequency has been widely studied in numerous contexts, because it rears its head in practically every situation involving the study of a quantum field in a non-trivial time dependent background (for a review, see for e.g.~\cite{txts1,txts2}). The dynamics of such a system is encoded in a propagator Kernel $K(q_2,t_2;q_1,t_1)$ which will allow us to determine the quantum state $\psi(q_2,t_2)$ at $t=t_2$ if the quantum state $\psi(q_1,t_1)$ at $t=t_1$ is given. Since the Lagrangian of the system is quadratic in $q$, the Kernel can be expressed in terms of the solutions to the classical equation of motion (see for e.g.~\cite{dr94}). So, if the classical solution for a harmonic oscillator with a particular time dependence is known, the quantum theory should be trivial.

And so one would have thought. The reason one continues to investigate this problem is not so much because the equations cannot be solved but because the interpretation of the solutions is non-trivial. (In this paper, for example, the solution appears within the first ten equations and the rest of the paper is on interpreting it!) Hence, it is worthwhile to raise some questions which need to be addressed in reasonably precise terms, before we proceed further. This will provide the motivation for contributing yet another paper on this topic to the already extensive literature!  

(1) In virtually every context we are interested in, there will be a classical degree of freedom symbolically denoted by a variable $C$ (cosmological background, external electromagnetic field, .....) interacting with a quantized degree of freedom $q$ (usually a scalar field) with the total Lagrangian for the combined system written as $L(q,C)=L_0(C)+L_I(C,q)$. We will be interested in quantizing $q$ in the background provided by $C$ and study the effect of $q$ on $C$ at the semiclassical level. To the lowest order, the configuration of $C$ will be determined by the equations of motion arising from $L_0$, ignoring $q$ completely. If this configuration is nontrivial (say, time dependent) then the quantum theory of $q$ will be based on a time dependent Hamiltonian and the $q$- particles will be generated by the interaction. The first conceptual question is: how can one define a notion of such particles and their production when the time dependence is nontrivial? If the classical system reaches a time-independent state asymptotically, it is straightforward to define the notion of particles in the \emph{asymptotic} \emph{in} and \emph{out} states and also obtain an expression for the \emph{total} number of particles produced. But many physically relevant problems (like cosmological particle production) will not give us this luxury of well defined in and out states. 

(2) The question of defining the notion of a particle in a time dependent background is not one of idle curiosity. If $C$ is producing the quanta of $q$, it has to supply the energy for the process and obviously this will modify the evolution of $C$. It will be important to obtain the equations of motion for $C$ with this backreaction incorporated as the particle production progresses. This is certainly not an asymptotic notion and one would like to imagine that -- in any causal theory -- the backreaction on $C$ at time $t=t_1$ should not depend on how the system will evolve at time $t>t_1$. So, purely conceptually, we need a notion of back reaction which does not use the concept of an out vacuum state etc.

(3) Such a back reaction --- and in fact, the notion of semiclassical evolution --- can be meaningful only if we have at least an approximate notion of `classicality' for the $q$ mode. Our intuitive idea of a particle which is produced (that has drained away energy from $C$) is classical and one assumes that a particle which is produced `stays produced'. This is of course not true in general and no sensible, time dependent, definition of particle exists which will obey this criterion. This is related to the fact that particle production is stochastic and what is usually computed is the {\it mean number} of particles, which is the mean of a stochastic process; if the variance is not small compared to the mean, one needs to review the entire philosophy.
On the other hand, we expect the notion of particles to be reasonably well defined if the quantum state is `quasi-classical'. This brings us to the question of defining the notion of classicality of the state in some precise sense and relating it to particle production.

The standard approach to understanding the evolution of the $C-q$ system involves starting with the path integral for the full system and integrating over the $q$ degree of freedom to obtain an {\it effective Lagrangian} $L_{eff}=L_0(C)+L_{corr}(C)$ in terms of $C$, that determines the transition amplitude between the asymptotic in and out vacuum states of the quantum subsystem \textit{when they are definable}. This effective Lagrangian, in general, can have an imaginary part, which 
can be related to the mean number of $q$-particles produced out of the vacuum by the $C$-field over the entire evolutionary history of the system. (Roughly speaking it specifies the out-particle content of the in-vacuum state.) In order to compute the {\it backreaction} of the quantum subsystem on the classical degree of freedom, one needs to vary $L_{eff}(C)$. However, since $L_{eff}$ can be complex, one would end up with complex equations (and solutions), which are difficult to interpret (see for e.g.~\cite{brown}). What is usually done is to consider only the real part of the effective action in the calculation of backreaction and work with $L_{0}(C)+\textrm{Re}L_{corr}(C)$. This term $\textrm{Re}L_{corr}(C)$ is normally associated with vacuum polarization, and gives a contribution to the backreaction even in the absence of particle creation. On the other hand the vacuum-persistence-probability
is directly related to the total number of particles produced and is determined by $\textrm{Im}L_{corr}(C)$. In such an approach it is not clear whether the back reaction due to the production of particles is incorporated in $\textrm{Re}L_{corr}(C)$.
The dropping of the imaginary part, which {\it also} carries information about the quantum state of $q$, needs to be justified and one would like to have a formalism which provides a unified picture of the evolution of the quantum system in the background of $C$.

We will now briefly connect up the above comments, presented abstractly in terms of a $C$ and $q$, in two specific contexts widely studied in the literature. The first one is pair production in a constant electric field and the second one quantum field theory in an inflationary universe.

In the case of the so called {\it Schwinger effect} in QED~\cite{schwinger}, we study the interaction of a quantized, charged scalar field $\phi$ with a constant background classical electric field {\bf E} (see refs.~\cite{itz,efield} for a small sample of recent work and reviews). Usually one computes the in-out transition amplitude by integrating over the $\phi$ variable in the path integral [say, using Schwinger's proper time formalism] to obtain the one-loop effective action (known as the Euler-Heisenberg effective action) for {\bf E}:
\beq
\langle 0_{+} | 0_{-} \rangle_{\bf E}   =  \int {\cal D} \phi~ e^{ i \int {\cal L}(\phi, {\bf A}) d^{4} x } ~\equiv~  e^{ i W_{eff} } ~\equiv~  e^{ i \int {\cal L}_{eff} d^{4} x } .
\eeq
The effective Lagrangian ${\cal L}_{eff}$ turns out to be complex, and its (renormalized) real and imaginary parts are given by
\beq
\textrm{Re} {\cal L}_{eff} = -\frac{1}{(4 \pi)^{2}}\int_{0}^{\infty} \frac{ds}{s^3} \cos m^{2} s \l[ \frac{qEs}{\sinh qEs} -1 + \frac{1}{6}q^{2}E^{2}s^{2} \r]
\eeq
and
\beq
\textrm{Im} {\cal L}_{eff} = \frac{(qE)^{2}}{16 \pi^{3}} \sum_{n=1}^{\infty} \frac{(-1)^{n+1}}{n^{2}} e^{- n \frac{\pi m^{2}}{qE}}. 
\eeq
 The real part possesses a Taylor expansion in $q^2$ and the lowest order correction to the classical Maxwell Lagrangian can be easily computed. The standard approach to computing the corrections to Maxwell's equations proceeds using {\it only} this real part; since it is analytic in $q^2$ one can obtain a modified action incorporating quantum corrections order-by-order. 
 On the other hand, the density of pairs produced by the electric field can be obtained directly from the imaginary part of the effective Lagrangian but is {\it non-analytic} in the coupling constant $q$. One would have expected the production of these particle pairs to drain the energy of the electric field and lead to a back reaction. It is not clear how an approach based on $\textrm{Re}{\cal L}_{eff}$ encodes this effect contained in $\textrm{Re}{\cal L}_{eff}$ and this needs to be clarified.

As a second example, consider the standard inflationary cosmology. The quantity usually computed in this context is not the particle number density but is the magnitude of the power spectrum~\cite{ps}. Since one uses it as a source in the perturbed Einstein equations, we are clearly moving into the domain of back-reaction. One normally computes the two-point correlation function of a quantized scalar field evolving in the inflationary background, and the power spectrum [the fourier transform in k-space] of the two-point function of the {\it field} is then used in calculating the fluctuations in density\cite{infl1,infl2,cosmo,paddy03}. This is different from computing an effective Lagrangian or from computing the number density of particles produced by the exponentially expanding universe (which is almost never attempted in the study of inflationary cosmology). The two approaches are thus technically different, and one needs to ascertain how the two-point function for the field is related to the particle content, and the extent to which the vacuum polarization part contributes to it. 

An added complication to this issue is that, although the early time evolution of the fluctuations [when the wavelength of a given perturbation is much smaller than the Hubble radius] is adiabatic and a choice of well-motivated initial conditions based on a natural definition of the vacuum state is possible, the late time phase [following Hubble exit, when the modes turn super-Hubble] is highly non-adiabatic and one can not identify a unique asymptotic out vacuum state. The notion of particles is ill-defined in such a dynamic background, and one needs to come up with appropriate variables that can help quantify the content of the quantum state of the perturbations in this background. Once again we need a generalization of our concepts from an out vacuum state to a time dependent situation.
 
Another closely related issue stems from the fact that although the perturbations are assumed to be generated as vacuum fluctuations of a quantized field, their late time evolution at super-Hubble scales is treated using classical notions, with the only reference to their quantum origin being in the choice of initial conditions. Clearly, one needs to understand the validity of this assumption, and figure out how a quantum state can turn effectively classical. 
This has generally been regarded as an interpretational issue, and the usually adopted viewpoint in the literature hinges on the concept of $\it{decoherence}$~\cite{decoh}, which basically provides a mechanism for suppression of quantum interference terms between various states of the system (a generic quantum property) by taking into account its unavoidable interaction with a suitably defined environment. Mathematically, one proceeds by splitting the system under consideration into a part that is built out of that set of physical variables one is interested in, and an `environment' that consists of the collection of those degrees of freedom that are in some sense inaccessible; the properties of the subsystem of interest can then be adequately described by a reduced density matrix that is obtained by starting with the full density matrix and tracing over the inaccessible degrees corresponding to the environment. Under suitable conditions, the reduced density matrix can be shown to reduce to a diagonal form (in a suitable basis) with the off-diagonal elements, representing quantum interference effects, getting suppressed. 

Although the formalism of decoherence can be invoked under fairly general conditions as real quantum systems seldom exist in isolation, ways have been suggested to do away with the quantum environment, and interpret classicality in terms of alternative notions. In the context of inflation, phase space correlations of the quantum variables have been analyzed using the Wigner distribution~\cite{infl2,wig,wig2}, and classicality interpreted in terms of peaking on the corresponding classical trajectory. The emergence of classicality has also been tied with the phenomenon of squeezing of the quantum state~\cite{infl2,class}. Because of the potential implications of such approaches, which rely only on the intrinsic properties of the system for an understanding of quantum-to-classical transitions, there is a need to make them unambiguous.  

Keeping these questions in mind, we will confine our attention to addressing two specific issues: (1) quantifying the physical content of a quantum state evolving in a time dependent background; in particular, identifying suitable variables that encode various aspects of the information contained in the quantum state, including the notion of particles, and (2) making the idea of interpreting classicality of the quantum state, as well as its relation with the concept of [appropriately defined] particle production, more precise.  

This will be done in a sequence of two papers. In the present paper, we will concern ourselves with studying the quantum mechanics of a single oscillator. This, of course, has general applicability. If one is studying a quantum field in an external time dependent background, like an FRW cosmological model \cite{parker} characterized by a scale factor $a(t)$ or an electric field expressed in a time dependent gauge with vector potential ${\bf A}(t)$, the field can be decomposed by taking a spatial fourier transform into a set of uncoupled oscillators with time dependent parameters, each corresponding to a particular fourier mode {\bf k}. This allows for a mode-by-mode analysis of the field evolution. Furthermore if, say, the oscillator labeled by the wave vector {\bf k} is in the $n$th excited state, this is interpreted, in the field picture, as the presence of $n$ particles with momenta {\bf k} each. Presuming such a close correspondence with the field picture, we will freely make use of field theory language in our dealings with the quantized oscillator, with the understanding that the various definitions are to be looked at in the broader context of fields. Having said that, it may also be mentioned that when one considers quantum fields, some additional ideas can emerge. Firstly, the presence of an infinite number of degrees of freedom brings in issues related to regularization, which need to be dealt with in a suitable manner. Secondly, one has the liberty of shifting from the oscillator picture by taking an inverse fourier transform back to coordinate space, and this opens up an alternative way to try understanding the physical content of the system. The case of quantum fields will be taken up in Paper II~\cite{gm}.

This paper is organized as follows. In section~\ref{sec:formalism}, we outline a well-motivated and straightforward formalism, in the Schrodinger picture, to analyze the dynamics of a general quantized time-dependent oscillator, that proves a convenient tool to understand the content of the evolving quantum state, and the relationships between the various physically sensible quantities built out of it. Then, in section~\ref{sec:analysis}, we first carry out a detailed analysis of a simple toy model, with particular emphasis on the question of relating particle content to classicality, and demonstrate that peaking of the Wigner function on the classical trajectory \emph{alone} is \emph{not} tantamount to having classical behavior; one needs to {\it also} look at some alternative measure of correlations to get rid of such ambiguities. We define such a reasonable alternative (calling it the `classicality parameter') and based on an approximate asymptotic analysis of a class of models with general time dependent frequency $\omega(t)$, show that {\it whenever} there is [appropriately defined] particle creation at late times, it is {\it always} accompanied by growth of the classicality parameter which can be quite unambiguously interpreted as an emergence of classicality. 
This is followed up with an analysis of a few additional toy examples, which serve to validate our analytic approximations and further illustrate the general features associated with the time evolution. Finally, we conclude in section~\ref{sec:diss} with a discussion. In what follows, we shall set $\hbar=c=1$.


\section{Evolution of the quantum state:~General formalism} \label{sec:formalism}

The starting point of our analysis is a general one-dimensional harmonic oscillator with the following action, where we treat the mass and the frequency of the oscillator to be time dependent:
\beq
{\cal A}[q] = \frac{1}{2} \int{m(t)\l[\dot q^2 - \omega ^2 (t)\, q^2\r]} dt \label{action}
\eeq
with the conditions that $m(t)>0,\omega^2(t)>0$. (If $m(t)$ is also monotonic, then using a time coordinate $\tau$ with $d\tau=dt/m(t)$ will get rid of $m(t)$ and change $\omega^2$ to $\omega^2 m^2$. But we will see that it is convenient not to do this and hence we will retain this form.) The classical trajectory of the oscillator can be obtained by solving the following equation of motion:
\beq
\frac{d}{dt} \left( m(t) \frac{dq}{dt} \right)+m(t) \omega^{2} (t) q = 0.  \label{cleq}
\eeq
As mentioned earlier, the action in \eq{action} can describe a particular fourier mode of a quantum field in the presence of a time-dependent classical background. In the standard procedure of canonical quantization, the variable $q$ describing the coordinate of the oscillator is replaced by the corresponding hermitian operator $\hat{q}$ satisfying the commutation relation $[\hat{q},\hat{p}]=i$.

In the Schrodinger picture which we shall work in, the evolution of the quantized oscillator can be described by a wave function $\psi (q,t)$ satisfying the time-dependent Schrodinger equation:
\beq
i\, \frac{\pa \psi(q,t)}{\pa t}
=-\frac{1}{2\, m(t)}\, \frac{\pa^2\psi(q,t)}{\pa q^2}
+\frac{1}{2}\, m(t) \omega^{2} (t) q^{2}\, \psi(q,t).\label{seq}
\eeq
We are interested in the solution to this equation which could have been interpreted as the ground state of the oscillator at some time $t=t_0$. This suggests looking for a particular class of solutions of \eq{seq} that has been studied in various contexts~\cite{gauss,pad,wig,paddy03,tpgswfn,ss}, the {\it form invariant} Gaussian state, which is an exponential of a quadratic function of $q$. If one restricts oneself to states which have a vanishing mean (which happen to be the relevant ones in the examples we will consider), then the wave function assumes the form
\beq
\psi\l(q,t\r)
= N(t)\, \exp\l[- R(t) q^2 \r]
\label{gswfn}
\eeq
where $N(t)$ and $R(t)$ are complex 
quantities. 
 Substituting the expression in \eq{gswfn} in the Schrodinger equation \eq{seq}, the following equations for $R$ and $N$ are obtained:
\beq
i\frac{\dot N}{N} = \frac{R}{m} \quad;\quad i \dot{R} = \frac{2 R^2}{m} -\frac{m \omega^{2}}{2}.
\label{nr}
\eeq
From the first equation and its complex conjugate, it is easy to show that
\beq
\vert N\vert^2=\l(\frac{R+R^\ast}{\pi}\r)^{1/2}
\label{eq:NkRk}
\eeq 
which can also be obtained from the normalization condition on the wave function.
Therefore, other than the overall phase of $N$, the only non trivial aspect of the quantum state is encoded in the time dependence of the function $R(t)$.

Before we present the exact solution to this system, it is worthwhile to obtain the adiabatic limit of the system. If $\omega,m$ are slowly varying functions of time, then the \eq{nr} can be solved to give 
\begin{equation}
R(t)\approx m(t)\omega(t)/2;\qquad 
N(t)=N_0\exp \l(-\frac{i}{2}\int_{t_0}^t\omega(t')dt'\r).
\label{adiaN}
\end{equation} 
This determines the evolution of our state in the adiabatic approximation.
Since $R(t)\approx m(t)\omega(t)/2$ in the adiabatic limit, it is convenient to introduce a new complex function $z(t)$ in place of $R(t)$ by the relation
\begin{equation}
R(t)\equiv\frac{m \omega}{2} \l(\frac{1-z}{1+z} \r). 
\end{equation} 
Clearly, $z$ measures the deviation of $R$ from the adiabatic value and we will call it the \emph{excitation parameter}.
The wave function is completely determined by $z(t)$ as
\beq
\psi (q,t) =  N \exp[- R q^{2}]  = N \exp \l[ - \frac{m \omega}{2} \l(\frac{1-z}{1+z} \r) q^{2} \r] = N \exp \l[ - \frac{m \omega}{2} \l(\frac{1-|z|^{2} - 2 i Im z}{1+|z|^{2} + 2 Re z} \r) q^{2} \r] \label{gswfn_z}
\eeq 
where 
\beq
|N|^{2} = \sqrt{\frac{m \omega}{\pi} \frac{\l(1-|z|^{2}\r)}{|1+z|^{2}}}.
\eeq 
 From the equation satisfied by $R$ in \eq{nr} one can obtain the equation satisfied by $z$; one finds that $z$ satisfies a rather simple first order differential equation:
\beq
\dot z + 2 i \omega z + \frac{1}{2} \l(\frac{\dot \omega}{\omega} + \frac{\dot m}{m} \r) (z^{2} - 1) = 0.  \label{eq:z_gen}
\eeq
Once this equation is solved for $z$ with some appropriate initial condition ($z=0$ set at some instant $t_{0}$) we have completely solved the problem. Our analysis will be based on the time evolution of $z$.

The evolution equation for $z$ can be written in a slightly different form by using $\omega$, instead of the time coordinate $t$ as the independent variable; given any monotonic range of $\omega(t)$, one can in principle invert this relation to obtain time as a function of the frequency, $t \equiv t(\omega)$. With this replacement, \eq{eq:z_gen} can be recast in the following form:
\beq
\omega \frac{d z(\omega)}{d \omega} + \frac{2 i}{\epsilon(\omega)}z(\omega) + \frac{1}{2}\l( z^2(\omega) - 1 \r)  =  0 \label{eq:z_omega}
\eeq  
with $\epsilon \equiv ((\dot{\omega}/\omega) + (\dot{m}/m))/\omega$, which will henceforth be called the \emph{adiabaticity parameter} (and which too is expressible as a function of $\omega$). This equation determines the function $z(\omega)$, and this conversion will be of particular relevance in our discussion of the effective Lagrangian in section~\ref{sec:L_eff}. 

It may be mentioned in passing that the adiabaticity parameter as defined above has a simple physical interpretation in the context of quantum fields in a cosmological background characterized by the scale factor $a(t)$. For a massless minimally coupled scalar field, any given fourier mode (describable by an oscillator) has a time dependent mass $a^3(t)$ and frequency $|{\bf k}|/a(t)$, so the adiabaticity parameter is $\epsilon_{\bf k}(t)= 2 \dot{a}(t)/ |{\bf k}|$, which (apart from a numerical factor) is just the ratio of the physical wavelength of the mode, $\lambda_p = 2 \pi a|{\bf k}|^{-1}$, to the Hubble radius $R_H=(\dot{a}/a)^{-1}$.  

The equations for $R,z$ are first order but nonlinear. Since they are of the generalized 
 Riccati type they can be transformed into second order linear equations. For example, if we set $R=-\l(i\, m/2\r)(\dot \mu/ \mu)$ where $\mu(t)$ is a new function then \eq{nr} implies that $\mu$ satisfies the following differential equation:
\beq
\ddot \mu+ \frac{\dot m}{m}\, \dot \mu
+\omega ^2 \mu=0
\label{mueq}
\eeq
which is same as the classical equation of motion, \eq{cleq}, satisfied by the oscillator variable $q$. The solutions of \eq{mueq} obey the Wronskian condition
\beq
\mu \dot \mu^{*} - \mu^{*} \dot \mu = -i\frac{W}{m(t)}
\eeq
where $W$ is independent of time. 
The variable $z$ can now be expressed as
\beq
z(t) =  \l( \frac{\omega + \frac{i}{\mu}\frac{d \mu }{dt}}{\omega -\frac{i}{\mu}\frac{d\mu}{dt}}\r).
\label{z}
\eeq
As mentioned in the introduction, we now have completely solved the problem.
 One can, in principle, solve \eq{mueq} to obtain the two linearly independent solutions for $\mu$, say $s$ and $s^{*}$. The general solution is a linear superposition of the form $\mu =\l[{\cal A}\, s + {\cal B}\, s^*\r]$, with ${\cal A}$ and ${\cal B}$ determined by one's choice of initial conditions. Once $\mu$ has been found, the function $R$ can then be computed, which completely fixes the quantum state of the system. The wave function $\psi(q,t)$ depends only on the ratio $\cal R = \cal B / \cal A$, since it is independent of the overall scaling of $\mu$. The real difficulty is not in doing this but in understanding what the physical content of the quantum state is. We shall now address this question.

\subsection{Particle content of the quantum state} \label{sec:pc}

The quantum state given by \eq{gswfn} is more generally known in the literature as a {\it squeezed} quantum state and has been quite extensively studied~\cite{squeeze} especially in the context of quantum optics. In the squeezed state formalism in the Heisenberg picture, one usually introduces the "squeeze" operator, defined as $\hat{S}(\xi) = \exp \l[(1/2)(\xi^{*} a^2 - \xi a^{\dagger 2})\r]$ with $\xi = r e^{i\theta}$, and where $r$ is known as the squeeze parameter ($0\leq r \leq \infty$). The function $z(t)$ that we defined earlier happens to be indirectly related to the squeeze parameters, as $z = - e^{i \theta - 2 i \rho} \tanh r$ (where $\dot{\rho}=\omega$). Unfortunately, the squeezed state formalism does not help much in our tasks in hand; so we shall follow a different approach here.

A physically motivated and reasonable way of quantifying the time-dependent content of the state would be to compare it with the \textit{instantaneous} energy eigenstates at any given moment. Since the oscillator parameters are time-dependent, one can not define stationary states in the usual manner; the alternative would be to define a set of instantaneous eigenstates at {\it every} instant. A wave function that starts off in the instantaneous ground state might, at a later time, be in a superposition of instantaneous eigenstates defined at $\it{that}$ moment; this can be thought of as excitation of quanta. 

We will define these instantaneous eigenstates, at a given moment $t$, as a set of states that have been obtained by {\it adiabatically} evolving the eigenstates at some initial instant $t_0$. This amounts to arranging matters in such a way that \textit{if} the oscillator evolves adiabatically, a wave function that has been set in the instantaneous ground state at $t=t_0$ will evolve to coincide, at every subsequent moment $t$, with the instantaneous ground state defined at $t$. This set of instantaneous eigenstates at time $t$ has the form
\beq
\phi_{n}(q,t) =  \l(\frac{m \omega}{\pi}\r)^{1/4}\frac{1}{\sqrt{2^{n} n!}} H_{n}(\sqrt{m \omega}q) \exp\left[-\frac{m \omega}{2} q^{2} - i\int_{t_0}^{t}\l( n + \frac{1}{2} \r) \omega(t) dt\right]
\label{inststate}
\eeq 
where $H_n$ are the Hermite polynomials~\cite{grad} and $n=0, 1, 2,...$. Except for the phase factor, these are essentially the harmonic oscillator wave functions, constructed using the instantaneous values of $\omega(t),m(t)$ at time $t$. They form a complete set of orthonormal functions at time $t$ and any other wave function can be expanded in terms of them. The phase in \eq{inststate} is so chosen that it coincides with the phase of the wave function in the adiabatic limit obtained in \eq{adiaN}. 

We can now expand our exact solution $\psi$ in terms of the instantaneous eigenfunctions. Since $\psi$ is an even function, the amplitude for the oscillator to be in the $nth$ instantaneous eigenstate $\phi_{n}(t)$ at time $t$ is non-zero only for even $n$, and is given by
\beq
C_{n}(t) = \int_{-\infty}^{\infty} {dq} \phi_{n}^{*}(q,t) \psi (q,t) = N \l(\frac{m \omega}{\pi}\r)^{1/4}\frac{1}{\sqrt{2^{n} n!}}\int_{-\infty}^{\infty} {dq} H_{n}(\sqrt{m \omega}q) e^{-\l(R+\frac{m \omega}{2}\r)q^{2} + i\int_{t_0}^{t}\l(n + \frac{1}{2} \r)\omega(t) dt}.   \label{amplitude}
\eeq
The above integral can be evaluated by standard methods (see Appendix A for details).
We can then compute the probability associated with the amplitude $C_{n}(t)$ to obtain
\beq
P_{2n}(t) = |C_{2n}|^{2}
= P \frac{(2n)!}{(n!)^2} \frac{|z|^{2n}}{2^{2n}},   \label{prob0}
\eeq
with the \emph{time-dependent} vacuum-persistence-probability $P_{0}(t)$ given by:
\begin{equation}
P_{0}(t)=P = \sqrt{1-|z|^2}
\label{defp0}
\end{equation} 
Clearly the probability for occupying the excited states
is controlled by $z$ justifying the name \emph{excitation parameter}. 
The generating function for the  probability distribution in \eq{prob0}, defined as $G(x) = \sum_{n=0}^{\infty} P_{2n} x^{n}$, is given by
\beq
G(x) = \frac{P}{\sqrt{1-x|z|^2}} = \sqrt{\frac{1-|z|^2}{1-x|z|^2}}. \label{genfn1}  
\eeq
Once the probability distribution is known, it is trivial to compute the mean number of quanta in the state at any time $t$; this is given by
\beq
\langle n \rangle = \sum_{n=0}^{\infty} 2n P_{2n} = 2\,G'(1) = \frac{P|z|^{2}}{(1-|z|^2)^{3/2}} = \frac{|z|^2}{1-|z|^2}.
\label{n}
\eeq
We shall take this quantity, $\langle n \rangle$, as describing the time dependent `particle' content of the quantum state. We do not claim that our definition is \emph{unique} in any sense; only that it will be \emph{useful} and physically reasonable. This interpretation will be borne out by different considerations in what follows and the first of these is the following. The system evolves under the action of a time-dependent Hamiltonian $H$ and one can compute the mean value of the energy at any time $t$ by the expectation value of the Hamiltonian, $E(t)=\langle \psi|H|\psi \rangle$. Direct computation shows that
\beq
E(t) 
=\l(\frac{m}{2\, W}\r)\l(\vert \dot \mu \vert^2 
+\omega ^2\, \vert \mu\vert^2\r)
=\l(\langle n \rangle +\frac{1}{2}\r)\, \omega (t).
\label{E-n}
\eeq
This clearly strengthens the motivation to think of $\langle n \rangle$ as defined above as the number of quanta present at time $t$. 

Let us rewrite the expression for the vacuum persistence probability, \eq{defp0} in terms of the the mean particle number:
\beq
P_{0}(t) = \sqrt{1-|z|^{2}} = \l( 1 + \langle n \rangle \r)^{-1/2} = \exp \l[ -\frac{1}{2}\ln \l( 1 + \langle n \rangle \r) \r].   \label{p0_n}
\eeq
It is obvious from this expression (as well as, of course, from \eq{prob0}) the excitation probability for different levels is \textit{not} Poissonian. When the excitation to level $2m$ is interpreted as creation of $m$ pairs of particles in quantum field theory, this implies that pair production is not --- in general --- a Poisson process and is non-trivially correlated. In the limit of $\langle n \rangle \ll1$, however, we have $P_{0}\sim\exp(-\langle n \rangle/2)$
which allows one to identify $\langle n \rangle/2$ as the mean number of pairs produced from the vacuum, which is a standard result.

It is clear from the expression for the generating function, \eq{genfn1}, that the mean particle number $\langle n \rangle$, as well as the higher order moments, are functions of only the magnitude of $z$. On the other hand, the wave function in (\ref{gswfn_z}) is built out of not only the magnitude, but also the phase $\theta$ of $z$. In fact, writing $z=|z|e^{i\theta}$ and using \eq{n} to determine $|z|$, we find that the wave function can be expressed in the form:
\begin{equation}
\psi(t,q)=
\frab{m\omega}{\pi}^{1/4}\left(1+2\langle n \rangle+2\sqrt{\langle n \rangle(\langle n \rangle+1)}\cos\theta\right)^{-1/4}
\exp\left[-\frac{1}{2}m\omega q^2\left(
\frac{1-2i\sqrt{\langle n \rangle(\langle n \rangle+1)}\sin\theta}{1+2\langle n \rangle+2\sqrt{\langle n \rangle(\langle n \rangle+1)}\cos\theta}
\right)\right].
\end{equation} 
It is obvious that the 
particle number only contains partial information about the quantum state; given the particle number (and even all the higher moments) at a given moment of time, one can {\it not} have complete knowledge of the state of the oscillator at that moment. 

While on this issue, it is worth noting that $\langle n \rangle\to\infty$ when $|z|\to1$. In this limit the width of the wave function scales as $\langle n \rangle$ and the gaussian spreads all over the coordinate space. In the same limit the width of the wave function in momentum space goes to zero, ``squeezing" the wave function to the $p=0$ axis. One can also show that the dynamical equations are same as the following equations for $\langle n \rangle$ and $\theta$:
\begin{equation}
\dot{\langle n \rangle}=\l(\frac{\dot\omega}{\omega} + \frac{\dot m}{m} \r)
\sqrt{\langle n \rangle(\langle n \rangle+1)}\cos\theta;\quad
\dot\theta= -2\omega-\frac{1}{2} \l(\frac{\dot\omega}{\omega} + \frac{\dot m}{m} \r)
\frac{2\langle n \rangle +1}{\sqrt{\langle n \rangle(\langle n \rangle+1)}}
\sin\theta.
\end{equation} 
It is obvious from these equations that even when $\dot\omega$ and $\dot m$ have a fixed sign, the sign of $\dot{\langle n \rangle}$ depends on the phase $\theta$ and hence need not be monotonic in general. Moreover, it may be noted that if $\langle n \rangle$ is specified as a function of time, one can, using the above equations, determine the form of $\theta(t)$, and this allows one to reconstruct the complete time evolution of the wave function. In this sense, it is possible to fix the state of the system completely given the time-variation of the mean particle number alone. And further, if $\langle n \rangle $ is a monotonic function of time, then one can trade off the $t$ dependence for dependence on $\langle n \rangle$, and thus express $z$ (as well as the wave function) explicitly in terms of just the mean number of particles. Such a transformation will be well-defined only when there is a one-to-one relation between $t$ and $\langle n \rangle$. (If the mean particle number is oscillatory, then there would in general be several values of time for the same $\langle n \rangle$, and consequently the expression for the wave function in terms of $\langle n \rangle$ will be multiple-valued as well, and not uniquely specifiable.) 

Getting back to our quantum state, it is clear that an additional physical variable needs to be specified that can provide information about the phase $\theta$. A suitable choice for this is the spread in the wave function, given by the expectation value of $q^{2}$. It has the form 
\beq
\langle q^{2} \rangle = \int_{-\infty}^{\infty} q^{2} | \psi(q,t) |^{2} d q = \frac{1}{2 ( R + R^{*} )} = \frac{|\mu|^{2}}{W}.
\eeq 
In the context of quantum fields in a cosmological setting, this dispersion can be directly related to the logarithmic power spectrum~\cite{infl1,infl2,pad,wig,cosmo}, which is the fourier transform of the two-point correlation function for the field evaluated in a particular state:
\beq
k^{3} P_{\phi} (k) = \frac{k^{3}}{2 \pi ^{2}} \langle q_{k}^{2} \rangle. 
\label{pow}
\eeq
The quantity $\langle q^{2} \rangle$ also can be re-expressed in terms of $z$ as follows:
\beq
\langle q^{2} \rangle = \frac{\langle n \rangle}{2 m \omega}  \l| 1+\frac{1}{z} \r| ^{2} = \frac{\l ( 2 \langle n \rangle + 1 \r)}{2 m \omega} + \frac{(\langle n \rangle +1)}{ m \omega} \textrm{Re}(z).   \label{q2_n}
\eeq
This expression, again, shows that the power in the $q$-mode is {\it not} completely expressible in terms of the instantaneous mean particle number [which involves just the magnitude of $z$], because it carries additional information encoded in the phase of $z$. It follows that, in general the computation of the power spectrum in \eq{pow} (in the cosmological context, for example) is different from computing the mean number of particles produced. However, it can be seen from \eq{q2_n} that in the limit when the real part of $z$ becomes a constant, $\langle q^{2} \rangle$ {\it can} be written purely as a function of the mean particle number.  

Our treatment is completely equivalent to the standard analysis that is generally done in the Heisenberg picture, in which the time dependence of the system is encoded in the Bogolyubov coefficients $\alpha(t)$ and $\beta(t)$ satisfying the relation $|\alpha(t)|^{2} - |\beta(t)|^{2} = 1$. The function $z$ that we have defined is related in a simple manner to these Bogolyubov coefficients, as
\beq
z(t) = \frac{\beta^*(t)}{\alpha^*(t)}~e^{- 2 i \rho(t)}   \label{H_pic}
\eeq
with $\dot{\rho}(t) \equiv \omega(t)$ (see Appendix C). In fact it can be shown quite trivially, that if one sets the oscillator in the vacuum state at time $t_0$, then the above quantity is directly proportional to the [normalized] amplitude for the vacuum state defined at time $t_0$ to be a 2-particle state with respect to the vacuum defined at time $t$:
\beq
\frac{\langle 2, t ~|~ 0, t_0 \rangle}{\langle 0, t | 0, t_0 \rangle} = \frac{1}{\sqrt 2}\frac{\langle 0, t ~|\hat{a}(t)\hat{a}(t)|~ 0, t_0 \rangle}{\langle 0, t | 0, t_0 \rangle}   = \frac{1}{\sqrt 2} \frac{\beta^*(t)}{\alpha^*(t)} \equiv \frac{z(t)}{\sqrt 2} e^{2 i \rho(t)}
\eeq
where $\hat{a}(t)$ is the annihilation operator defined at time $t$, with $\hat{a}(t) |0,t \rangle = 0$.  
 It follows from the relation in \eq{H_pic} that all the physical quantities are alternatively expressible in terms of the Bogolyubov coefficients. In particular \eq{n} now reduces to $\langle n \rangle=|\beta|^2$ which is the standard result. We however shall not be using them in the analysis that follows, but will stick to understanding the evolution of $z(t)$. (In field theory, each fourier mode labeled by a wave vector ${\bf k}$ will have a corresponding $z_{\bf k}$ and one can also obtain the spatial fourier transform of this quantity; we will see in Paper II~\cite{gm} that it contains valuable information about the classicality of the state.)

\subsection{Effective Lagrangian}  \label{sec:L_eff}

Another useful way to quantify the effects induced by the time-dependent background on the quantum system, especially in the semiclassical context, is to look at the {\it vacuum persistence amplitude}, which measures the amplitude for a state to be a vacuum at late times, if it started out as a vacuum state at early times. Normally, this quantity is defined using asymptotic in and out vacuum states~\cite{txts1,txts2,schwinger}. It is directly related to the effective action, the imaginary part of which specifies the asymptotic particle content in the quantum state, while the real part is used in analyzing backreaction. 

This idea needs generalization when asymptotically adiabatic vacuum states cannot be defined. One would like to have some sort of a generalized time-dependent analogue of the effective action that is amenable to a suitable interpretation. Based on the formalism outlined in the previous section, one can define such a quantity in a fairly natural manner. 
 
 Consider a situation in which the oscillator has been set in the instantaneous ground state at some instant, say, $t=t_{0}$. This fixes the form of $z(t)$ and hence of $R(t)$. One can then compute the amplitude for this oscillator to evolve and be in the \textit{instantaneous} ground state at some moment $t>t_0$ in the future: this is just the amplitude $C_{0}$ evaluated in \eq{amplitude}, and one can write it in terms of a time dependent `effective action' as follows:
\beq
C_{0}(t) = \frac{ N(t) \l( m \omega \pi \r)^{1/4} }{ \sqrt{\l( R(t) + \frac{m \omega}{2} \r)} } \exp\left[i\int_{t_0}^{t}\frac{\omega(t)}{2}dt\right] \equiv \exp\{i A_{eff} (t)\} \equiv \exp\left[i \int_{t_{0}}^{t} L_{eff}(t) dt\right].
\eeq  
Since the amplitude is known it can be easily computed and turns out to be a rather simple expression:
\beq
L_{eff}(t) = \frac{i}{4} \l( \frac{\dot \omega}{\omega} + \frac{\dot m}{m} \r) z  \label{L_eff}
\eeq
To place this result in context, recall that the time dependence of $\omega$, for example, arises because the classical degree of freedom (background metric, electric field .....) is described by a time dependent solution. This makes quantities like $\dot\omega=(\partial\omega/\partial C)\dot C$ explicit functionals of $C$. Further, the effective Lagrangian also depends on $C(t)$ implicitly through $z$ which is determined in terms of the background variables through the differential equation \eq{eq:z_omega}. While the effective Lagrangian looks rather simple when expressed in the above form this simplicity is deceptive (and useless for the purpose of calculating backreaction etc.). The effective Lagrangian needs to the thought of as a \emph{functional} of background variables and this is often nontrivial. Nevertheless, we can make significant progress using the above expressions, keeping the relevant caveats in mind.

The real and imaginary parts of the effective Lagrangian are directly related to the function z. 
The real part can be written as
\beq
\textrm{Re} L_{eff}(t) = - \frac{1}{4} \l( \frac{\dot \omega}{\omega} + \frac{\dot m}{m} \r) \textrm{Im} z =  -\frac{1}{4}~\epsilon~\omega~\textrm{Im} z.  \label{RL_eff} 
\eeq 
Note that we have defined the particle content and the vacuum state using the instantaneous eigenstates obtained by adiabatic evolution. Therefore, the usual adiabatic term (integral of $\omega$ over time) does not appear in this expression and $Re L_{eff}$ vanishes in the adiabatic limit ($\epsilon\to0$). 
The imaginary part, on the other hand, is given by
\beq
\textrm{Im} L_{eff}(t) = \frac{1}{4} \l( \frac{\dot \omega}{\omega} + \frac{\dot m}{m} \r) \textrm{Re} z.   \label{IL_eff1}
\eeq
We would have expected it to be related to the particle content and indeed it is. Using the equation for $\langle \dot n \rangle$, one can show that this is equal to
\beq
\textrm{Im} L_{eff}(t) = \frac{1}{4} \frac{d}{dt} \ln ( 1 + \langle n \rangle );
 \quad \textrm{Im} A_{eff}=\frac{1}{4}\ln ( 1 + \langle n \rangle )
 \label{IL_eff}
\eeq
This is in accordance with the standard interpretation of $A_{eff}$ if the system has a late time adiabatic regime, where $\textrm{Im} A_{eff}(t)$ saturates to a constant value specifying the asymptotic particle number. If $\langle n \rangle\ll1$ then the imaginary part of the effective action becomes $\textrm{Im} A_{eff}(t)\approx (1/4)\langle n \rangle$ so that the vacuum persistence probability is 
\begin{equation}
|\langle 0,t|0,t_0 \rangle |^2\approx \exp(-2\textrm{Im} A_{eff}(t))\approx 1-\frac{1}{2}\langle n \rangle
\end{equation} 
which matches with  the result in \eq{defp0} for $|z|^2\ll 1$. (The factor $(1/2)$ is due the fact that $\langle n \rangle$ is the mean number of \textit{particles} while in quantum field theory one usually quotes the result for mean number of \textit{pairs}.) We will see that our definitions make sense in all other cases too when we look at the behavior of $L_{eff}$ in different limits in the study of toy models in the next section.
 
To recapitulate, we have defined the quantum state in terms of a parameter $z$ and find that the particle content of the state determines (and is determined by) $|z|^2$. To fix the state uniquely we also need to know the phase of $z$, which can be obtained from some other suitably defined quantity like the dispersion $\langle q^2 \rangle$.
Together they completely specify the state of the system at any moment. 
The next question we want to address is the `classicality' of this state, and its approach to classicality. This may be understood, in one possible way, by shifting attention to the system's phase space.
  
\subsection{Wigner function}

We now will attempt to quantify the level of classicality of the quantum state in terms of its phase space correlations. A suitable tool for this purpose is the Wigner distribution function \cite{wig,wig2,pad}, which is defined, for a wave function $\psi(q,t)$, as
\beq
{\cal W}\l(q, p,t\r)
=\frac{1}{2\pi}\, \int\limits_{-\infty}^{\infty}du\;
\psi^*\l(q+\frac{u}{2},t\r)\; 
\psi\l(q-\frac{u}{2},t\r)\; 
e^{ipu}.
\eeq
 The Wigner function can be regarded as a quantum analogue of the classical distribution function, and satisfies the following identities:
\beq
\int_{-\infty}^{\infty} {\cal W}(q,p) dp = |\psi(q)|^{2},
\eeq
\beq
\int_{-\infty}^{\infty}  {\cal W}(q,p) dq = |\varphi(p)|^{2}
\eeq
where $\varphi(p)$ is the wave function in Fourier space (i.e, the fourier transform of $\psi$). These two relations suggest that $ {\cal W}(q,p)$ can be thought of as a probability distribution in phase space, provided it is positive.
In general, Wigner function satisfies the evolution equation:
\beq
\frac{\pa {\cal W}}{\pa t} + \frac{p}{m}\frac{\pa {\cal W}}{\pa q}-\frac{dV}{dq}\frac{\pa {\cal W}}{\pa p} = \frac{\hbar ^2}{24}\frac{d^{3}V}{dq^{3}}\frac{d^{3}{\cal W}}{dp^{3}}+ ...     \label{wfee}
\eeq
where $V(q)$ denotes the potential, and dots indicate terms with higher powers of $\hbar$ and higher derivatives of $\cal W$ and $V$. For quadratic potentials, the right hand side of eq.(\ref{wfee}) vanishes and we recover the classical continuity equation. (This is true for any potential  up to order $\hbar^{2}$.) Although the above relations suggest that the Wigner function might be interpreted as a `joint probability distribution', it can take on negative values in some cases~\cite{squeeze}. Gaussian states, however, turn out to have a positive-definite Wigner function.

The Wigner function is useful in studying quantum to classical transitions in a system in terms of correlations between the phase space variables $(q,p)$; a pure quantum system is represented by a completely uncorrelated Wigner function like 
 ${\cal W}(q,p)=A(q)B(p)$ so that the probability for the system to have a momentum $p$ is independent of its position $q$. The ground state of the harmonic oscillator, for example, has such a Wigner function showing it is very non-classical. For a classical system, we will find ${\cal W}$ to be peaked in a limited region in phase space with the totally classical state being ${\cal W}(q,p) \propto \delta (p-f(q))$, where $p=f(q)$ represents the classical phase-space trajectory. Hence the classical system is expected to show a strong correlation between $q$ and $p$. A useful measure of the `classicality' of a state can thus be provided by the degree of correlation between $q$ and $p$. For this purpose, we consider the following quantity:
\beq
{\cal S} = \frac{\langle pq \rangle_{{\cal W}}}{\sqrt{\langle p^2 \rangle_{{\cal W}} \langle q^2 \rangle_{{\cal W}} }}
\eeq
with the average for any function $F(q,p)$ of the phase space variables being calculated using the Wigner function:
\beq
\langle F(q,p) \rangle_{{\cal W}} = \int_{-\infty}^{\infty} \int_{-\infty}^{\infty} F(q,p) {\cal W}(q,p) dq dp.
\eeq
We shall henceforth refer to the object ${\cal S}$ as the {\it classicality parameter}; for a pure quantum state, like the ground sate of the oscillator, we get ${\cal S} = 0$. For a highly classical state, we will expect ${\cal S}\to 1$.

The Wigner function corresponding to the gaussian wave function in \eq{gswfn} can be expressed as
\beq
{\cal W}\l(q, p, t\r)
=\frac{1}{\pi}\; 
\exp\l[-\frac{q^2}{\sigma^2(t)}
-\sigma ^2(t)\, \l(p-
{\cal J}(t)\,  q\r)^2\r]  \label{gswifn}
\eeq
where $\sigma $ and ${\cal J}$ are given by
\beq
\sigma ^2=\l(R+R^{\ast}\r)^{-1}
\qquad{\rm and}\qquad {\cal J}= i\,\l(R-R^{\ast}\r).
\eeq
The above expressions can be written in terms of the function $z$ (and $\langle n \rangle$) as follows:
\beq
\sigma ^2 = \frac{2\vert \mu \vert ^{2}}{W} = \frac{|1+z|^2}{m \omega (1-|z|^2)} = \frac{\langle n \rangle}{m \omega} \l |1+\frac{1}{z} \r|^2
\label{sigma}
\eeq
and
\beq
{\cal J} = \frac{m}{2} \frac{d (ln\vert \mu \vert ^{2})}{dt} = \frac{2 m \omega Im(z)}{|1+z|^2}.
\label{J}
\eeq
For this Wigner distribution, the classicality parameter is expressible in terms of $\sigma ^2$ and ${\cal J}$:
\beq
{\cal S} = \frac{{\cal J} \sigma ^2 }{\sqrt{1 + ({\cal J} \sigma ^2 )^{2}} }.
\eeq
It can be seen from eq.(\ref{gswifn}) that when ${\cal J}=0$, the Wigner function represents an uncorrelated product of gaussians in $q$ and $p$, and ${\cal S}=0$; this happens to occur when the gaussian state coincides with an instantaneous ground state. On the other hand, when both $\sigma ^2$ and ${\cal J}$ take on non-zero values, ${\cal W}(q,p)$ becomes correlated and ${\cal S}$ would be appreciably different from zero. [It may be noted that $|{\cal S}|\leq 1$, and the maximum possible value that it can have is unity.] The choice of ${\cal S}$ as a measure of classicality is thus a reasonable one.

The idea of associating particle creation with the approach to classicality of a quantum state has been around in the literature for quite some time, particularly in the context of cosmology~\cite{class,wig}. It has been demonstrated, in the case of perturbations evolving during an inflationary epoch in the early universe, that the Wigner function gets peaked on the corresponding classical trajectory as the state undergoes large squeezing at super-Hubble scales~\cite{pad,wig}. (The argument in this case hinges on the fact that the frequency of the oscillator turns imaginary at late times, and results on peaking of the Wigner function for the inverted oscillator are then applied to draw the relevant conclusions.) Here, we are interested in exploring the generality of this connection. In particular, we would like to know how production of particles (identified, in our case, with a growth in the quantity $\langle n \rangle$) is related to the spreading of the Wigner function, as well as to the variation in the sharpness of the $q$-$p$ correlation, i.e. the classicality parameter.

We note that for the time dependent oscillator (with completely general $m(t)$ and $\omega(t)$), both the particle number and the $q$-$p$ correlation can be built out of the complex quantity $z$; but there is no simple relation \emph{directly} connecting the two. The classicality parameter involves phase information of $z$ as well, and this makes drawing conclusions about its behavior solely on the basis of one's ideas about $\langle n \rangle$ (which depends only on the magnitude of $z$) not so straightforward. This means that one needs to actually determine $z$ to address the question of relating the two variables, to which we now turn.  
	

\section{Analytic approximations and asymptotic analysis}  \label{sec:analysis}

We move on to explicit analysis of several concrete examples based on the ideas of section~\ref{sec:formalism}. For convenience, we set $m=1$, so that the time dependence enters only through $\omega(t)$. (As we explained earlier, this entails no loss of generality.) The appropriate measure to characterize the behavior of the oscillator is, of course, the adiabaticity parameter given by $\epsilon(t)=\dot{\omega}/\omega^{2}$, whose magnitude determines the nature of its dynamics. We will assume the late time limit to be given by the time variable $t$ going to infinity, and consider the two extreme cases: one in which the late time evolution is adiabatic ($\epsilon \ll$ 1) and the other in which the adiabaticity condition is strongly violated ($\epsilon \gg 1$). 

For the case of an oscillator with a time-dependent frequency $\omega(t)$, the equation (\ref{eq:z_gen}) for the function $z$,  simplifies to
\beq
\dot z + 2 i \omega z + \frac{\dot \omega}{2 \omega} (z^{2} - 1) = 0.  \label{eq:z}
\eeq
As mentioned before, once $z$ is known, all other variables of interest can be trivially obtained. The table given below encapsulates the expressions for the various physical quantities in terms of $z$ for ready reference: 
\vspace{0.1cm}
\begin{center}
  {\scriptsize
\begin{tabular}{|c|c|}
  \hline
 & \\
  &  $\psi (q,t) = \exp[- R q^{2}]$ \\
The wave function & \\
 & = $\exp \l[ - \frac{\omega}{2} \l(\frac{1-z}{1+z} \r) q^{2} \r] $ \\
& \\
\hline
   & \\
~~~~~~~~Evolution equation for z~~~~~~~~&~~~~$\dot z + 2 i \omega z + \frac{\dot \omega}{2 \omega} (z^{2} - 1) = 0$~~~~\\
&    \\
\hline
&  \\
Mean particle number & $\langle n \rangle = \frac{|z|^2}{1-|z|^2}$   \\
&  \\
\hline 
&  \\
 Wigner function ${\cal W}\l(q, p, t\r)$ &  $\sigma ^2 = \frac{|1+z|^2}{\omega (1-|z|^2)}$   \\
 &  \\
~~~$=\frac{1}{\pi}\; 
\exp\l[-\frac{q^2}{\sigma^2(t)}
-\sigma ^2(t)\, \l(p-
{\cal J}(t)\,  q\r)^2\r] $~~~   &  ${\cal J} = \frac{2 \omega Im(z)}{|1+z|^2}$\\
& \\
\hline
& \\
Classicality parameter &   $ {\cal J} \sigma ^{2}  = 2 \langle pq \rangle_{{\cal W}} $ \\
& \\
${\cal S} = \frac{{\cal J} \sigma ^2 }{\sqrt{1 + ({\cal J} \sigma ^2 )^{2}} }$  &  $= \frac{2 Im(z)}{(1-|z|^{2})}$\\
& \\
\hline
& \\
Spread in the wave function  &  $\langle q^{2} \rangle = \frac{\langle n \rangle}{2 \omega}  \l| 1+\frac{1}{z} \r| ^{2}$ \\
& \\
(related to the power spectrum) &  = $\frac{\l ( 2 \langle n \rangle + 1 \r)}{2 \omega} + \frac{(\langle n \rangle +1)}{ \omega}Re z$ \\
& \\
\hline
& \\
& $Re L_{eff} = -\frac{1}{4} \frac{\dot \omega}{\omega} Im z$ \\
Effective Lagrangian  &  \\ 
& ~~$Im L_{eff} = \frac{1}{4} \frac{\dot \omega}{\omega} Re z = \frac{1}{4} \frac{d}{dt}\ln(1+\langle n \rangle)$~~ \\
& \\
\hline

\end{tabular}
}
\end{center}
\vspace{0.3cm}

We are particularly interested in how an oscillator, that starts off in the instantaneous ground state at some moment $t=t_{0}$, would evolve at late times [the $t \to \infty$ limit]. With this in mind, we will specifically consider a scenario in which the system enters an adiabatic phase at late times and, alternatively, when the asymptotic evolution deviates from adiabaticity. All the useful choices for the frequency $\omega (t)$ fall within one of these categories. Before we move on to carry out an approximate analysis for such functions, we will take up a simple toy model first. 

\subsection{The case of constant adiabaticity parameter}  \label{sec:ce}

We will begin our analysis by considering the case of {\it constant} adiabaticity parameter $\epsilon$. (In the cosmological context, this model would describe a massless scalar field evolving in a background characterized by the scale factor $a(t) \propto t$, which corresponds to having a matter source with equation of state $p=-\rho/3$. This is the borderline case between accelerating and decelerating models.) This model is exactly solvable, and can describe adiabatic as well as non-adiabatic evolution through appropriate choices for the value of $\epsilon$. This flexibility allows one to see how the nature of the evolution changes with transition from adiabaticity to non-adiabaticity. This example is also expected to shed some light on the general features one may expect to find in the two extreme limits.

For the case of constant $\epsilon$, the frequency function is given by
\beq
\omega(t) = \frac{\lambda}{1 + \nu (\lambda t)} \qquad(0<t<\infty)
\eeq
with $\nu > 0$, $\lambda$ having the dimension of frequency and the adiabaticity parameter being equal to $-\nu$. The combination $(\nu \lambda)^{-1}$ sets the time scale for variation; for $t \gg (\nu \lambda)^{-1}$, the frequency falls as $1/(\nu t)$.
For this $\omega(t)$, equation (\ref{eq:z}) for $z$ can be solved analytically to obtain the following result (where $\tau \equiv \lambda t$ and we have set $z=0$ at $t=0$):
\beq
z(\tau) = \frac{1 - (1 + \nu \tau)^{\frac{z_{1}-z_{2}}{2}}}{z_{1}(1 + \nu \tau)^{\frac{z_{1}-z_{2}}{2}} - z_{2}} 
\eeq
with $z_{1,2} = (2i \pm \sqrt{\nu^{2} - 4})/\nu $.

\begin{figure}
\begin{center}
\subfigure[ ]{\label{clen} \includegraphics[width=6cm,angle=0.0]{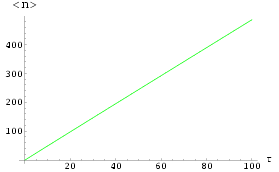}}
\subfigure[ ]{\label{cleS}\includegraphics[width=6cm,angle=0.0]{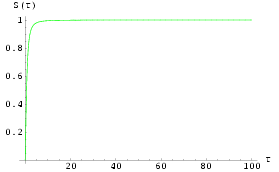}}
\subfigure[ ]{\label{clez}\includegraphics[width=5cm,angle=0.0]{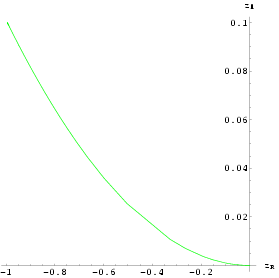}}
\end{center}
\caption{Variation of the mean particle number $\langle n \rangle$, classicality parameter ${\cal S}$ and the excitation parameter $z$ with $\tau$ for constant $|\epsilon|= 20$. $\langle n \rangle$ is a monotonically increasing function of time. ${\cal S}$ increases sharply and saturates at unity at large times. The trajectory of $z$ remains confined within a quadrant and ends up at a point with $|z|=1$ in the $\tau \to \infty$ limit.}
\label{cle}
\end{figure}

\begin{figure}
\includegraphics[width=6cm,angle=0.0]{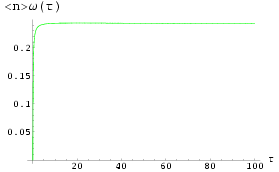}
\caption{Variation of $\langle n \rangle \omega$ with time for the case of $|\epsilon|= 20$. This quantity saturates asymptotically, so the energy remains finite at late times.}
\label{cleno}
\end{figure}

It is convenient to analyze the behavior of this model by splitting the range of $\nu$ into two parts: one corresponding to $\nu^2>4$ and the other to $\nu^2<4$. Let us consider the former case first.
For $\nu^2>4$, the expression for the particle number is given by
\beq
\langle n \rangle = \frac{\nu^2}{(\nu^2 -4 )} \sinh^2 \l(\frac{1}{2} \sqrt{1-\frac{4}{\nu^{2}}} \ln(1+\nu \tau) \r).
\eeq
This is an exact expression valid at all times. It is clear that the particle number increases monotonically with time in this case. Since we are particularly interested in the scenario of highly non-adiabatic evolution, it is appropriate to consider large values of $\nu$ (in comparison with unity). For $\nu \gg 1$, we have the following approximate expression for the particle number:
\beq
\langle n \rangle  \approx \frac{1}{4} \frac{\nu^2 \tau^2}{1+ \nu \tau}.
\eeq 
Let us consider the early and late time limits of the above expression. For the case of $\nu \tau \ll 1$,
\beq
\langle n \rangle \approx \frac{\nu^2 \tau^2}{4} + O(\nu^3 \tau^3).   \label{n_Leps_early}
\eeq
On the other hand, for $\nu \tau \gg 1$,
\beq
\langle n \rangle \approx \frac{ \nu \tau}{4}  \approx \frac{1}{4 \omega}.  \label{n_Leps_late}
\eeq 
For strongly non-adiabatic evolution, starting from a vacuum state, the particle number increases without bound at late times. But \eq{n_Leps_late} also shows that the energy content of the particles $\langle n \rangle \omega$, however, saturates at a constant value (1/4) at late times. Our approximation is borne out by the plots of the particle number and the quantity $ \langle n \rangle \omega$, shown in figures~\ref{clen} and \ref{cleno}. The particle number grows without bound, while the latter quantity has a constant limiting value. 

The classicality parameter ${\cal S}$ for this case, describing the strength of phase space correlations, is plotted in figure~\ref{cleS}. It is evident that ${\cal S}$ starts from zero [corresponding to the initial vacuum state], and quickly grows to unity as the particle number increases with the progress of time. This clearly presents an example where the phase space correlations grow in accompaniment to particle creation due to violation of adiabaticity. 

One can also directly visualize the trajectory of the state in the complex $z$ plane, which has been plotted in figure~\ref{clez}. $z$ starts from the origin, corresponding to the initial vacuum state, stays within a quadrant, and ends up at a limiting point corresponding to $|z|=1$ at late times.

\begin{figure}
\subfigure[ ]{\includegraphics[width=6cm,angle=0.0]{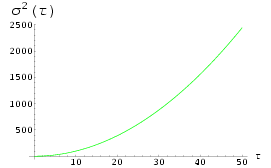}}
\subfigure[ ]{\includegraphics[width=6cm,angle=0.0]{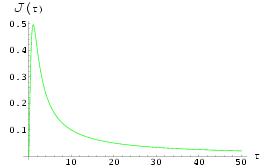}}
\caption{Variation of the functions $\sigma^2$ and ${\cal J}$ for $|\epsilon|= 20$. $\sigma^2$ grows monotonically while ${\cal J}$ falls to zero at late times, indicative of a Wigner function that peaks on the $q-$axis.}
\label{cleW}
\end{figure}

One would also like to know how the Wigner function evolves with time. This can be understood by computing the functions $\sigma^2$ and ${\cal J}$ using the relations (\ref{sigma}) and (\ref{J}). These functions have been plotted in figure~\ref{cleW}. The value of $\sigma^2$ starts from $1/\omega(t=0)$ and monotonically increases with time, while ${\cal J}$ starts from zero, rises to a maximum, and then falls off again to zero at late times. This variation corresponds to a Wigner function that ends up peaking on the $q-$axis at late times. Let us compare this with the way the corresponding classical trajectory behaves with time. The classical trajectory is given by
\beq
q_c (\tau)  =  \sqrt{1 + \nu \tau} \l( c_1 e^{\frac{1}{2}\sqrt{1-\frac{4}{\nu^2}} \ln(1 + \nu \tau)} + c_2 e^{-\frac{1}{2}\sqrt{1-\frac{4}{\nu^2}} \ln(1 + \nu \tau)} \r)
\eeq
and
\beq
p_c(\tau) = q'_c(\tau) = \frac{\nu}{2} \l[ \l(1+\sqrt{1-\frac{4}{\nu^2}}\r)c_1 \l( 1 + \nu \tau \r)^{\frac{1}{2}\l(\sqrt{1-\frac{4}{\nu^2}}-1\r)} + \l(1-\sqrt{1-\frac{4}{\nu^2}}\r)c_2 \l( 1 + \nu \tau \r)^{-\frac{1}{2}\l(\sqrt{1-\frac{4}{\nu^2}}+1\r)}  \r] 
\eeq 
where $c_1,c_2$ are real constants. It is clear that in the late time limit, $q_c \to \infty$ while the momentum $p_c \to 0$, so every trajectory at sufficiently late times ends up on the $q-$axis, coinciding with the peaking of the Wigner function. This feature, together with the behavior of the classicality parameter, provides strong indication of a quantum-to-classical transition taking place at late times.

The numerical value of the effective Lagrangian for this case can be evaluated using \eq{L_eff}. (As we mentioned before, such an expression has a limited validity since the effective action should be treated as a functional of $\omega$ rather than as a function of $t$ for a particular $\omega(t)$. Nevertheless it illustrates certain interesting features.) In the highly non-adiabatic limit of $\nu \gg 1$, it has the approximate form
\beq
L_{eff} \approx  \frac{\lambda}{1 + \nu \tau}\l[ \frac{\nu^2 \tau^2}{2 (\nu \tau + 2 )^2} + \frac{i \nu}{4} \frac{\nu \tau}{ (\nu \tau + 2 )} \r].
\eeq
This reduces to the following limiting form at early times [corresponding to $\nu \tau \ll 1$]:
\beq
L_{eff} \approx \omega \l[ \frac{\nu^2 \tau^2}{8} + \frac{i \nu}{8} \nu \tau  \r].   \label{RL_Leps}
\eeq
Interestingly,  the real part of the effective Lagrangian can be re-expressed in terms of the particle number using \eq{n_Leps_early}:
\begin{equation}
\textrm{Re} L_{eff} \approx \langle n \rangle \omega / 2
\end{equation} 
This shows that the real part of the effective Lagrangian does contain information about the particle production  (and hence can encode it in the back reaction calculation) in the non-adiabatic, but $\langle n \rangle \ll 1$ limit. We will find later that this is indeed a general feature.
On the other hand, in the late time limit,
\beq
L_{eff} \approx \lambda \l[ \frac{1}{2 \nu \tau} + \frac{i \nu}{4} \frac{1}{\nu \tau}  \r] \approx \frac{\lambda}{8} \l( \frac{1}{\langle n \rangle} + \frac{i}{2}\frac{\nu}{\langle n \rangle}  \r)~~\text{for}~ \nu \tau \gg 1.
\eeq
From the above expression, it is clear that although $L_{eff}(\tau) \to 0$ in the late time limit, the effective {\it action} is logarithmically divergent, since $L_{eff} \propto \tau^{-1}$. This is related to particle production proceeding without bound at late times.

\begin{figure}
\subfigure[ ]{\label{csen}\includegraphics[width=6cm,angle=0.0]{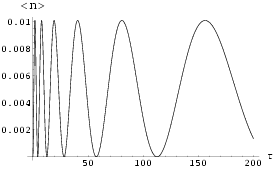}}
\subfigure[ ]{\label{cseS}\includegraphics[width=6cm,angle=0.0]{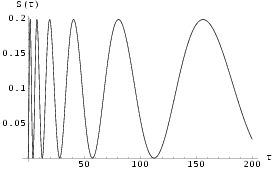}}
\subfigure[ ]{\label{csez}\includegraphics[width=5cm,angle=0.0]{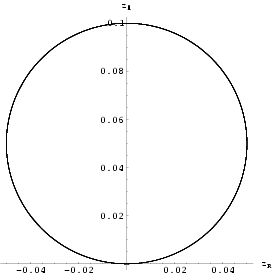}}
\caption{Variation of the mean particle number, classicality parameter and $z$ for constant $|\epsilon|= \nu = 0.2$. Both$\langle n \rangle$ and ${\cal S}$ are oscillatory at all times with constant finite amplitudes. The complex trajectory of $z$ is a circle in the upper half-plane, reflected in the oscillatory nature of $\langle n \rangle$ and ${\cal S}$.}
\label{cse}
\end{figure}

We next consider the other regime, covering values of the adiabaticity parameter lying in the range $\nu^{2} < 4$. For this case, $z_1,z_2$ are purely imaginary and the expression for $z$ may be rewritten as
\beq
z = \frac{1 - e^{i \sqrt{\frac{4}{\nu^{2}}-1} \ln(1+\nu \tau)}}{z_{1}e^{i \sqrt{\frac{4}{\nu^{2}}-1} \ln(1+\nu \tau)} - z_{2}}.
\eeq 
The particle number has the exact form
\beq
\langle n \rangle = \frac{\nu^2}{(4-\nu^2)} \sin^2 \l[\frac{1}{2} \sqrt{\frac{4}{\nu^{2}}-1} \ln(1+\nu \tau) \r].
\eeq
This quantity keeps oscillating between the values 0 and $\nu^2 / (4 - \nu^{2})$, representing a continuous interplay of creation and annihilation of quanta. This situation thus presents a case where the oscillations persist with constant amplitude and the particle number never settles down to a constant value [or enters a phase of monotonic variation]. We have plotted the variation of the particle number as well as the classicality parameter for this case [with $\nu$ chosen to be 0.2] in figures~\ref{csen} and \ref{cseS}. The classicality parameter is also oscillatory in the absence of a monotonic variation in $\langle n \rangle$, and this may be contrasted with the non-adiabatic case considered earlier, where ${\cal S}$ grows and saturates at unity. This is suggestive of the possible close connection between particle creation and growth of phase space correlations, which will be explored in greater depth in the examples that will follow.   

The complex trajectory of $z(t)$ is also depicted in figure~\ref{csez}. The values of the real and complex parts of $z$ oscillate with time, and the trajectory is a circle with the center displaced from the origin. The magnitude of $z$, too, oscillates and this is reflected in the variation of the particle number. This evolution differs sharply from that for large adiabaticity parameter, where the number of particles as well as the $q$-$p$ correlation display monotonic increase with time.

\begin{figure}
\subfigure[ ]{\includegraphics[width=6cm,angle=0.0]{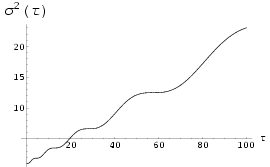}}
\subfigure[ ]{\includegraphics[width=6cm,angle=0.0]{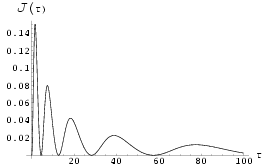}}
\caption{Variation of the quantities $\sigma^2$ and ${\cal J}$ for $|\epsilon|= 0.2$. $\sigma^2$ is an increasing function of time while ${\cal J} \to 0$ asymptotically, so the Wigner function ends up peaking on the $q-$axis.}
\label{cseW}
\end{figure}

As for the Wigner function, its evolution is described by the functions $\sigma^2$ and ${\cal J}$, shown in figure~\ref{cseW}. While ${\cal J}$ goes to zero at late times, $\sigma^2$ is an increasing function of time, driving the Wigner function to get concentrated on the $q-$axis at late times. In comparison, the classical trajectory has the general form
\beq
q_c(\tau) = c_1 \sqrt{1 + \nu \tau } \cos \l( \frac{1}{2} \sqrt{\frac{4}{\nu^{2}}-1} \ln( 1 + \nu \tau) + \phi \r) 
\eeq
where $c_1$ is real and $\phi$ is a constant phase factor, and
\beq
p_c(\tau) = q'_c(\tau) = \frac{\nu c_1}{2 \sqrt{1 + \nu \tau }} \l[ \cos \l( \frac{1}{2} \sqrt{\frac{4}{\nu^{2}}-1} \ln( 1 + \nu \tau) + \phi \r) - \sqrt{\frac{4}{\nu^{2}}-1}~ \sin  \l( \frac{1}{2} \sqrt{\frac{4}{\nu^{2}}-1} \ln( 1 + \nu \tau) + \phi \r)  \r].
\eeq

The amplitude of the momentum steadily decreases, while that of $q_c$ grows with time. The trajectory, consequently, ends up on the $q-$axis asymptotically. This matches with the late time behavior of the Wigner function. Thus the Wigner function peaks on the classical trajectory at late times, similar to that found in the non-adiabatic case earlier, in spite of the fact that here there is no genuine particle creation. It may also be noted that neither does the correlation measure ${\cal S}$ grow monotonically at large times. This example thus brings out a potential problem with regarding peaking of the Wigner distribution on the classical trajectory as the sole indicator of classicality. At the same time, the behavior of the classicality parameter, which evidently tracks particle creation, and is markedly different in the two cases, suggests that this measure may be able to provide a useful means of resolving such ambiguities regarding the interpretation of classicality.        

In the strongly adiabatic limit, we have $\nu \ll 1$, and the particle number is approximately given by
\beq
\langle n \rangle  ~\approx~ \frac{\nu^2}{4} \sin^2 \l(\frac{1}{\nu}\ln(1+\nu \tau) \r) + O(\nu^4).   \label{n_approx_Seps}
\eeq 
For small times ($\tau \ll 1$), $\ln(1+\nu\tau)\approx \nu \tau$ and $\sin \tau \approx \tau$, so that
\beq
\langle n \rangle  ~\approx~  \frac{\nu^2 \sin^2{\tau}}{4} ~\sim~ \frac{\nu^2 \tau^2}{4}.
\eeq
This behavior matches with the early time variation of $\langle n \rangle$ in the extreme non-adiabatic case (corresponding to $\nu \gg 1$).

The effective Lagrangian for $\nu^2 < 4$ takes the following form:
\beq
\textrm{Im} L_{eff} = \frac{1}{4}\frac{\dot{\langle n \rangle}}{\l(\langle n \rangle + 1 \r)} =  \frac{\omega}{2} \sqrt{1-\frac{\nu^2}{4}} \l[ \frac{\sin \l( \sqrt{\frac{4}{\nu^{2}}-1} \ln(1+\nu \tau) \r) }{ \frac{8}{\nu^{2}}- 1 - \cos \l( \sqrt{\frac{4}{\nu^{2}}-1} \ln(1+\nu \tau) \r) } \r]
\eeq
and
\beq
\textrm{Re} L_{eff} = \frac{\omega \nu^2}{8} \l[ \frac{\sin^{2} \l(\frac{1}{2}\sqrt{\frac{4}{\nu^{2}}-1} \ln(1+\nu \tau) \r) }{ 1-\frac{\nu^2}{8}\l( 1 + \cos \l( \sqrt{\frac{4}{\nu^{2}}-1} \ln(1+\nu \tau) \r) \r) } \r].
\eeq
In the adiabatic limit of $\nu \ll 1$, the real part is approximately given by
\beq
\textrm{Re} L_{eff} ~\approx~ \frac{\omega \nu^2}{8} \frac{4 \langle n \rangle }{\nu^2} \l[ 1 + \frac{\nu^2}{8}\l( 1 + \cos \l( \sqrt{\frac{4}{\nu^{2}}-1} \ln(1+\nu \tau) \r) \r) \r]  ~\equiv~ \frac{\omega \langle n \rangle}{2} \l[ 1 + O(\nu^2) \r]   \label{RL_Seps}
\eeq
and this is, of course, valid at all times. The identical form for $\textrm{Re}L_{eff}$ in both the limits of small as well as large values of the adiabaticity parameter, which may be noted from Eqs.(\ref{RL_Leps}) and (\ref{RL_Seps}), suggests that when the particle number is small, the real part of the effective Lagrangian is directly related to the particle content in the quantum state; this is  of  relevance to the issue of accounting for back-reaction due to particle creation using $\textrm{Re}L_{eff}$, a point we shall return to later. 

To summarize, the model of constant adiabaticity parameter, for different choices for the magnitude of $\epsilon$, can exemplify both adiabatic and non-adiabatic evolution. In the adiabatic limit of small $|\epsilon|$, particle creation is suppressed with the mean particle number and the classicality parameter (both starting from zero corresponding to an initial vacuum state) remaining finite and oscillatory at all times. This behavior sharply differs from that in the extreme non-adiabatic case of large $|\epsilon|$, where the mean number of particles as well as the $q$-$p$ correlation diverge with time, although the energy saturates at a finite limiting value. In both cases, however, the Wigner function ends up peaking on the corresponding classical phase space trajectory at late times, in marked contrast to the behavior of the classicality parameter.

Having gained some idea of the kind of features which may be expected in adiabatic as well as non-adiabatic evolution, we now move on to consider the case of a general adiabaticity parameter $\epsilon(t)$, and examine its two possible late time limits.
  
\subsection{ Adiabatic evolution at late times }  \label{sec:adiab}

For a general $\omega(t)$ we can physically distinguish the following situations. Depending on the numerical value of $\epsilon(t)$ one can describe the evolution as adiabatic or non-adiabatic. Further, in the case of adiabatic evolution one can further distinguish between two cases. the first case corresponds to the system evolving in an arbitrary fashion till about $t=T$, say, and evolving adiabatically for $t>T$. In this case the quantum state will not an instantaneous vacuum state at $t=T$ and certain amount of particle production would have taken place by then. The further adiabatic evolution should preserve this particle content with a higher order contribution to small amount of further particle production. The second case corresponds to an adiabatic evolution all the way from $t_0$ to $t$ with the system starting at the vacuum state at time $t=t_0$. In this case, the particle production will be a higher order effect and we will be interested in the lowest nontrivial order contribution.
We will first analyze the case of the evolution proceeding adiabatically at late times. This covers all possible choices for the function $\omega(t)$ which fall off slower than $1/t$ in the $t \to \infty$ limit, and correspondingly, $\epsilon(t) \to 0$. 

A perturbation theory approach can be adopted to find an approximate solution in an adiabatic regime. We start by writing the function $f(t)$ in terms of two new functions ${\cal A}(t)$ and ${\cal B}(t)$ as
\beq
f(t) = \frac{{\cal A}(t)}{\sqrt{2 \omega}} e^{i\theta(t)} + \frac{{\cal B}(t)}{\sqrt{2 \omega}}e^{-i\theta(t)} \qquad (\dot{\theta} \equiv \omega)  
\eeq
and imposing the additional constraint condition (to take into account our introduction of {\it two} new functions in place of the original one) that 
\beq
\dot{f}(t) = i \omega \frac{{\cal A}(t)}{\sqrt{2 \omega}} e^{i\theta(t)} - i \omega \frac{{\cal B}(t)}{\sqrt{2 \omega}}e^{-i\theta(t)}. 
\eeq
This gives (details can be found in the Appendix C)
\beq
\frac{\dot{f}}{f} = i \omega \l( \frac{{\cal A}(t)-{\cal B}(t)e^{-2i\theta(t)}}{{\cal A}(t)+{\cal B}(t)e^{-2i\theta(t)}} \r)
\eeq
and yields the following equations for ${\cal A},{\cal B}$:
\beq
\dot{\cal A} = \frac{\dot \omega}{2 \omega} {\cal B} e^{-2i\theta(t)} \quad,\quad \dot{\cal B} = \frac{\dot \omega}{2 \omega} {\cal A} e^{2i\theta(t)}.     \label{AB_eqs}
\eeq
The first order adiabatic approximation for the functions ${\cal A}(t),{\cal B}(t)$ can be found by assuming initial conditions ${\cal A}(t_0)=a,{\cal B}(t_0)=b$ (with $a,b$ being arbitrary complex constants) at some initial time $t_0$ within the adiabatic regime. To a first approximation, the coupled equations then give
\br
{\cal A}(t) ~&\approx&~ a + i\frac{b}{4} \l(\epsilon(t) e^{-2i\theta(t)} - \epsilon(t_0) \r) - \frac{ib}{4}\int_{t_0}^{t} d t' \dot{\epsilon}(t') e^{-2i\theta(t')}, \nonumber \\
{\cal B}(t) ~&\approx&~ b - i\frac{a}{4} \l( \epsilon(t) e^{2i\theta(t)} - \epsilon(t_0)  \r) + \frac{i a}{4}\int_{t_0}^{t} d t' \dot{\epsilon}(t') e^{2i\theta(t')}. 
\er
At the lowest order, one can ignore the integrals over $\dot{\epsilon}$ on the assumption of adiabaticity. (This result, therefore is exact for the the $\epsilon=$ constant case discussed earlier.) So, at the first order of approximation, we have
\br
z(t) ~=~ \frac{{\cal B}(t)}{{\cal A}(t)}e^{-2i\theta(t)} ~&\approx&~ \frac{b}{a}e^{-2i\theta(t)} + \frac{i}{4} \l( \epsilon(t_0) \l( 1 + \frac{b^2}{a^2} \r) e^{-2i\theta(t)} - \epsilon(t) \l( 1 + \frac{b^2}{a^2} e^{-4i\theta(t)} \r) \r) + O(\epsilon^2)  \nonumber \\
&\equiv& \frac{b}{a} e^{-2i\theta(t)} + {\cal R}(t) + O(\epsilon^2).    \label{z_approx}
\er
where the function ${\cal R}(t)$ is of order $\epsilon$. This solution is valid in the period when the evolution is adiabatic, irrespective of the past history of the system. For example, one can envisage an evolution in which the system starts at the ground state, evolves very non-adiabatically for a period of time, $t_i<t<T$ say, and is adiabatic for $t>T$. In that case, the system will be in a highly excited state (with some amount of particle production already having taken place during $t_i<t<T$) at $t=T$ and during $t>T$, this state will evolve adiabatically. In such a case we cannot say anything about the magnitude of $a,b$ and we will keep them arbitrary for the moment. (This situation should be contrasted with another one frequently discussed in the literature in which it is assumed that the evolution is adiabatic \emph{throughout} the period. We will consider this case later on.)
The corresponding $|z|^2$, to order $\epsilon$, is given by
\beq
|z|^{2} ~\approx~ \l|\frac{b}{a} \r|^2 - \frac{1}{2} Im \l[ \frac{b^*}{a^*} \l( \epsilon(t_0) \l( 1 + \frac{b^2}{a^2} \r) - \epsilon(t) e^{2 i \theta(t)} \l( 1 + \frac{b^2}{a^2} e^{-4i\theta(t)} \r) \r) \r] + O(\epsilon^2).   \label{z2_approx}
\eeq
Using this expression in \eq{n}, one can compute an approximation for the mean particle number, which, again to order $\epsilon$, is given by
\beq
 \langle n \rangle ~\approx~ \frac{|b|^2}{|a|^2 - |b|^2} - \frac{|a|^4}{2( |a|^2 - |b|^2 )^2} Im \l[ \frac{b^*}{a^*} \l( \epsilon(t_0) \l( 1 + \frac{b^2}{a^2} \r) - \epsilon(t) e^{2 i \theta(t)} \l( 1 + \frac{b^2}{a^2} e^{-4i\theta(t)} \r) \r) \r]  + O(\epsilon^2).
\eeq
Let us now consider various limits of this expression. 

If the system evolves very non-adiabatically for a period of time $t_i<t<T$  and is adiabatic for $t>T$ then we can take the $\epsilon(t)\to0$ limit at $t>T$.
Then, for any non-zero value of $b/a$, the terms dependent on $\epsilon(t)$ in the above expression get progressively suppressed. This implies that the mean particle number has a finite limiting value given by
\beq
\lim_{t \to \infty} \langle n \rangle ~\approx~ \frac{|b|^2}{|a|^2 - |b|^2} - \frac{\epsilon(t_0)}{2} \frac{|a|^4}{( |a|^2 - |b|^2 )^2} Im \l[ \frac{b^*}{a^*} \l( 1 + \frac{b^2}{a^2} \r) \r] + O(\epsilon^2 (t_0)).
\eeq
This result is completely understandable. In this case, all the particle production takes place during $t_i<t<T$ and the constants ${\cal A},{\cal B}$ play the role of Bogolyubov coefficients for the evolution in Heisenberg picture; the mean number of particles is indeed $|{\cal B}|^2$. 
If one compares the expression for $z$ in \eq{z_approx} with \eq{H_pic} relating $z$ to the functions $\alpha$ and $\beta$ in the Heisenberg picture, it can be seen that in the $\epsilon(t) \to 0$ limit, the ratio of the Bogolyubov coefficients becomes a constant: $\beta^* / \alpha^* \approx (b/a) + (i \epsilon(t_0)/4) ( 1 + (b^2/a^2) ) $. Using the Wronskian condition $|\alpha|^{2} - |\beta|^{2} = 1$, the limiting expression for $\langle n \rangle$ derived above can then be shown to be equal to $|\beta|^{2}$. This is exactly the expression one expects to obtain for the asymptotic particle number in an adiabatic `out' vacuum state in the Heisenberg picture. This thus demonstrates the close correspondence between our approach and the conventional one.
 
An expression for the $q$-$p$ correlation can also be worked out, using the approximations in \eq{z_approx} and \eq{z2_approx}, to first order in $\epsilon$:
\beq
{\cal J}\sigma^{2} ~=~ \frac{2 Im(z)}{(1-|z|^{2})} ~\approx~ 2 \l[\frac{ Im (a^* b e^{-2 i \theta})}{|a|^2 - |b|^2} + \frac{|a|^2 Im{\cal R}}{|a|^2 - |b|^2}  + \frac{ 2 Im \l( a^* b e^{-2 i \theta} \r) Re\l( a b^* {\cal R}e^{2 i \theta}\r)}{(|a|^2 - |b|^2)^2}\r] + O(\epsilon^2).
\eeq 
In the late time limit (as $\epsilon(t) \to 0$), the function ${\cal R}$ has a constant amplitude, and since $\theta$ is an increasing function of time [owing to our choice for $\omega(t)$], the above expression is oscillatory with a finite amplitude. Thus, in the late time adiabatic regime, once the particle number has saturated, the phase space correlation $\langle qp \rangle_{{\cal W}}$  remains bounded and oscillates around zero, again indicative of the necessity of \textit{continuing} particle excitation for driving the system towards classicality.

One can also look at the effective Lagrangian, defined in \eq{L_eff}, in this limit. The imaginary part of the action will become a constant at late times with its value related to the asymptotic number of particles produced, in conformity  with the usual interpretation. The real part, using \eq{RL_eff} and \eq{z_approx}, is given by
\br
\textrm{Re} L_{eff} &=& -\frac{1}{4}\frac{\dot \omega}{\omega}Im z \nonumber \\
&\approx& -\frac{\omega(t) \epsilon(t)}{4} Im\l(\frac{b}{a}e^{-2i\theta(t)}\r) - \frac{\omega(t) \epsilon(t)}{16} Re\l(\epsilon(t_0) \l( 1 + \frac{b^2}{a^2} \r) e^{-2i\theta(t)} - \epsilon(t) -  \epsilon(t) \frac{b^2}{a^2} e^{-4i\theta(t)} \r)  \label{RL_adia}
\er
to second order in the adiabaticity parameter. Written in this form, it is evident that the first O($\epsilon$) term is a rapidly oscillating function. The O($\epsilon^{2}$) term also contains two oscillatory functions. These average out to zero, and the first non trivial contribution to the real part of the effective action comes from the non-oscillatory term with magnitude $\omega \epsilon^{2}/16$ in \eq{RL_adia}.

Let us next consider the case in which the entire evolution up to some time $t$ is adiabatic, the oscillator being started off in the instantaneous vacuum state at some time $t_0<t$ in the adiabatic regime.

The initial conditions for an instantaneous vacuum correspond to setting $a=1,b=0$ and $\dot{f}/f=i\omega(t_0)$. For these initial conditions, because of \eq{AB_eqs}, even at the lowest order of approximation in the adiabatic regime ${\cal B}$ cannot be assumed to remain constant. To a first approximation, we now have:
\begin{equation}
{\cal A}(t) \approx 1 ~,~\dot{\cal B}(t) \approx \frac{\dot \omega}{2 \omega}e^{2i\theta(t)}
\end{equation} 
so that
\beq
{\cal B}(t) = -\frac{i}{4} \l( \epsilon(t) e^{2i\theta(t)} - \epsilon(t_0)  \r) + \frac{i}{4}\int_{t_0}^{t} d t' \dot{\epsilon}(t') e^{2i\theta(t')}.
\eeq
The integral over $\dot{\epsilon}$ will be again ignored under the assumption that $\dot\epsilon(t)\approx0$. So, at the lowest order of approximation, 
\br
z(t) = \frac{{\cal B}(t)}{{\cal A}(t)} e^{-2i\theta(t)} = -\frac{i}{4} \l( \epsilon(t) - \epsilon(t_0)  e^{-2i\theta(t)} \r), \nonumber \\
|z(t)|^{2}  = \frac{1}{16} \l[ \epsilon^{2}(t) + \epsilon^{2}(t_0) - 2 \epsilon(t)\epsilon(t_0)\cos \l( 2 \theta(t) \r) \r].  \label{z_adia_vac}
\er
This can also be derived from the equation for $z$:
\beq
\dot z + 2 i \omega z = \frac{\dot \omega}{2 \omega} (1 - z^{2}).
\eeq
Starting with $z(t_0)=0$ and assuming that to the lowest order of approximation $(1-z^2) \approx 1$ the above equation can be solved to obtain the same expression for $z(t)$ as found above. The approximate expression for $z$ gives
\beq
\frac{1}{1-|z(t)|^{2}}  \approx  1 + \frac{1}{16} \l[ \epsilon^{2}(t) + \epsilon^{2}(t_0) - 2 \epsilon(t)\epsilon(t_0)\cos \l( 2 \theta(t) \r) \r]  +   {\cal O}(\epsilon^{4}),
\eeq
so that the particle number to order $\epsilon^2$ is given by
\beq
\langle n \rangle \approx \frac{1}{16} \l[ \epsilon^{2}(t) + \epsilon^{2}(t_0) - 2 \epsilon(t)\epsilon(t_0)\cos \l( 2 \theta(t) \r) \r].   \label{n_ad_vac}
\eeq
If $\epsilon(t_0)=0$, then the particle number, to the lowest order in $\epsilon$, is just $\epsilon^{2}(t)/16$.

In order to verify the validity of our approximation, let us reconsider as an illustration, the example of constant adiabaticity parameter we analyzed earlier. For this case, $\epsilon(t) = -\nu$ and $\theta(t) = (1/\nu)\ln(1+ \nu \tau)$. [The $\dot{\epsilon}$ dependent integrals are exactly zero here.] This gives the approximate particle number from \eq{n_ad_vac} to order $\epsilon^2$ as
\beq
\langle n \rangle  \approx \frac{\nu^2}{8} \l[ 1 - \cos \l( \frac{2}{\nu} \ln(1+\nu \tau) \r) \r].  \label{con_eps_approx}
\eeq
This is clearly in agreement with the approximate expression derived in \eq{n_approx_Seps} from the exact analytic formula. 

It is also straightforward to compute the functions $\sigma^2$ and ${\cal J}$ using the approximation for $z$ in \eq{z_adia_vac}, and to ${\cal O}(\epsilon)$ are given by
\beq
\sigma^2 = \frac{1}{\omega(t)}\l( 1 + \frac{\epsilon(t_0)}{2}\sin(2\theta(t)) \r)
\eeq 
and
\beq
{\cal J} = -\frac{\omega(t)}{2} \l( \epsilon(t) - \epsilon(t_0) \cos(2\theta(t)) \r).
\eeq
The expressions for $\sigma^2$ and ${\cal J}$ make it clear that their late time behavior depends on the asymptotic limit of the frequency function; in particular, if $\omega(t) \to 0$ at large times, we have $\sigma^2 \to \infty$ and ${\cal J} \to 0$, so the Wigner function will peak on the $q$-axis (this happens for example in the constant $\epsilon$ case for $\nu \ll 1$). On the other hand, if $\omega(t)\to \infty$ (with $\epsilon(t) \to 0$), then $\sigma^2 \to 0 $ while ${\cal J}$ is oscillatory with an increasing amplitude at late times, representing a Wigner function that ends up peaking on the $p$-axis. In contrast, the $q p$ correlation, which has the form
\beq
\langle q p \rangle_{\cal W} = {\cal J}\sigma^{2} \approx -\frac{1}{2} \l( \epsilon(t) - \epsilon(t_0) \cos(2\theta(t)) \r) + {\cal O}(\epsilon^2)
\eeq 
is oscillatory in general and remains finite, so the classicality parameter, too, is bounded at late times.
   
For the adiabatically evolved vacuum state, the imaginary part of the effective Lagrangian again leads to the standard result and gives the effective number of particles produced. As for the real part, making use of \eq{z_adia_vac} one has, up to order $\epsilon^2$: 
\beq
\textrm{Re} L_{eff} ~\approx~ -\frac{1}{4}\frac{\dot \omega}{\omega} Im \l[ -\frac{i}{4} \l( \epsilon(t) - \epsilon(t_0)  e^{-2i\theta(t)} \r)  \r]  ~=~ \frac{\omega}{16} \l[ \epsilon^2 (t) - \epsilon(t) \epsilon(t_0) \cos (2 \theta(t)) \r].
\eeq 
Using \eq{n_ad_vac}, one can rewrite the above expression in terms of the mean particle number, to order $\epsilon^{2}$, as
\beq
\textrm{Re} L_{eff} ~\approx~  \omega \langle n \rangle  - \frac{\omega}{16} \epsilon^2 (t_0) + \frac{\omega}{16} \epsilon(t) \epsilon(t_0) \cos (2 \theta(t)). 
\eeq

The first term in this expression has a simple interpretation. The energy drained from the classical system due to particle production, $-\omega \langle n \rangle$, matches with this term. (Recall that effective Lagrangian and effective Hamiltonian differ by a sign in this case.) So to lowest order in the adiabaticity parameter, one can indeed incorporate the back reaction due to the production of particles using the first term in $\textrm{Re} L_{eff}$. The other two terms will be subdominant when $\epsilon(t_0)$ is sufficiently small and will represent transient effects. In this context, to lowest non vanishing order in $\epsilon$, the real part of the effective Lagrangian incorporates information about particle creation. 

\subsection{Non-adiabatic evolution at late times}  \label{sec:nonadiab}

We would like to compare the features suggested by the above analysis with the alternative scenario of adiabaticity being {\it violated} in the late time limit. All choices for $\omega(t)$ which monotonically fall off faster than $1/t$ for large $t$ would pass into such a phase, and for such functions, the adiabaticity parameter grows without bound at late times: $\lim_{t \to \infty} \epsilon(t) = \infty$.
 
Since our focus is on understanding the influence of particle creation on the evolution towards classicality, the case of increasing particle number $\langle n \rangle$ is considered. The approximate {\it asymptotic} form of the solution which has this feature, is given (see Appendix D for an outline of the derivation) to the lowest order by 
\beq
z = ( 1 + \Delta ) e^{i \theta},\nonumber 
\eeq
with
\beq
\Delta ~\sim~ c_{1} \omega \nonumber  ~,\qquad  \theta ~\sim~ \pi - c_{0} \omega - 2 \omega t  \label{delta_theta}
\eeq
in which $c_{0},c_{1}$ are real constants. 
With these analytic expressions, one can determine the asymptotic form of the mean particle number as well as the correlation:
\beq
\langle n \rangle \approx  - \frac{1}{ 2 c_{1} \omega}  \label{approx_n}
\eeq
and
\beq
2 \langle qp \rangle_{\cal W} ~=~ {\cal J} \sigma^{2} ~\simeq~ - \frac{\sin \theta}{\Delta}~ \approx~ -\frac{c_{0}}{c_{1}} - \frac{2 t }{c_{1}}.    \label{approx_S}
\eeq
Since it has been assumed that $\omega(t) \to 0$ at late times, the particle number clearly diverges. Further, $\langle qp \rangle_{\cal W}$ is an increasing function of time in the $t \to \infty$ limit. Thus, steady particle creation (i.e. growth in the particle number $\langle n \rangle$) at late times is accompanied by a monotonic growth of the magnitude of the $q$-$p$ correlation, and so the classicality parameter ${\cal S}$ ends up saturating at unity. 

For this case, the late time approximations for the imaginary and real parts of the effective Lagrangian, making use of \eq{delta_theta}, are given by
\beq
\textrm{Im} L_{eff} \approx \frac{1}{4} \frac{\dot{\langle n \rangle}}{n}  \sim  -\frac{1}{4} \frac{\dot \omega}{\omega}\qquad \textrm{Re} L_{eff} \approx - \frac{1}{2} \dot{\omega} t 
\eeq
from which it is evident that the imaginary part of the effective {\it action} blows up as rapid particle production occurs. In comparison, the real part of $L_{eff}$ falls to zero because of our constraint on the asymptotic form of $\omega(t)$, and $\textrm{Re} A_{eff}(t)$ has a finite asymptotic limit.  
\newline

Based on the features revealed by the above two cases, one can begin to delineate a relationship between the production of particles (increasing $\langle n \rangle$) and emergence of classical correlations in terms of the classicality parameter. We would also like to see how the corresponding Wigner function behaves in these two limits. In order to proceed with this, as well as to set the above ideas on a firmer footing, we shall next analyze two toy examples which will serve to demonstrate the key aspects. We will resort to a numerical approach for tracking the time evolution since it suffices for our purpose, which is to compare the two alternative late time scenarios and verify the validity of our approximations. These examples will also serve as prototypes for more realistic models, like those appearing in field theory in cosmological or electromagnetic backgrounds, to be discussed in Paper II~\cite{gm}.  
Let us begin with an example that provides an illustration of case~\ref{sec:adiab} considered above.
  
\subsection*{Example 1: Adiabatic evolution at late times}

We choose for the time dependent frequency the form $\omega_{1}(t) = \lambda \sqrt{1+\lambda^{2} t^{2}}$ ($-\infty < t < \infty$). This frequency function is symmetric in time $t$. It is physically relevant too, as it appears in the study of a complex scalar field evolving in the background of a constant, classical electric field (see for eg. Refs.~\cite{itz,efield}).

For this function, the adiabaticity parameter given by $\epsilon(t)= \lambda t/(1+\lambda^{2} t^{2})^{3/2}$ vanishes in the asymptotic limits ($t \to \pm \infty$), so in- and out- vacuum states can be defined in the adiabatic regions. If one starts out in the in- vacuum state (defined by $\langle n \rangle \rightarrow 0$ as $t \to -\infty$), the non-equivalence of the two asymptotically defined vacua implies that the state will appear populated with quanta measured with respect to the out- vacuum at late times. 

\begin{figure}[h]
\begin{center}
\subfigure[ ]{\label{o1n}\includegraphics[width=6cm,angle=0.0]{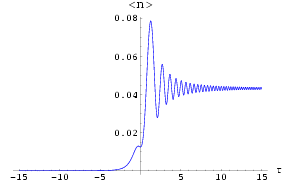}}
\subfigure[ ]{\label{o1S}\includegraphics[width=6cm,angle=0.0]{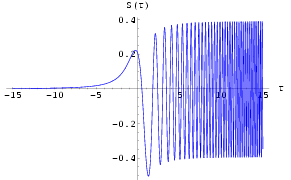}}
\subfigure[ ]{\label{o1z}\includegraphics[scale=0.5]{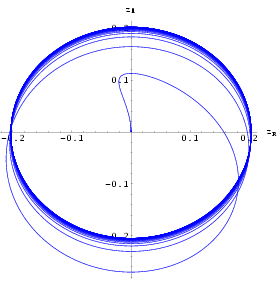}}
\end{center}
\caption{Plots of the particle number $\langle n \rangle$, classicality parameter ${\cal S}$ and the complex trajectory of $z$ with time for $\omega_{1}(t)$ ($\epsilon \to 0$ at late times). $\langle n \rangle$ saturates at a finite value at late times, and ${\cal S}$ ends up oscillating about zero with a finite amplitude. The variation of ${\cal S}$ reflects the complex trajectory of $z$, which at late times is a circle centered on the origin of the complex plane.}
\label{o1}
\end{figure}

The evolution of the system can be conveniently followed in our Schrodinger picture formalism. The solution to the equation for $z$, \eq{eq:z}, determines the complete time evolution of the quantum state (with the initial condition $z=0$ set well within the early time adiabatic phase at some moment $|t_{0}| >> \lambda^{-1}$ with $t_{0} < 0$). The numerical solutions for the mean particle number and the classicality parameter as functions of the dimensionless variable $\tau = \lambda t$ are plotted in figures~\ref{o1n} and \ref{o1S}. 

As anticipated, the particle number $\langle n \rangle$ starts from zero in the distant past, and saturates at a constant value in the $\tau \to \infty$ limit. However, in the intermediate region (where the adiabaticity parameter $\epsilon$ is appreciably non-zero), $\langle n \rangle$ is characterized by large osillations superimposed on a steadily increasing mean, which can be interpreted as a continual interplay of creation and annihilation of quanta. The amplitude of these oscillations gets progressively diminished as the evolution proceeds into the late time adiabatic regime, and the mean settles at a constant value. The oscillations in $\langle n \rangle$ conform to the results we have obtained in earlier sections using analytic approximations.

The classicality parameter ${\cal S}$ stays very close to zero at early times, but in the late time adiabatic regime, ends up oscillating about ${\cal S}=0$ as our approximate analysis in section~\ref{sec:adiab} did suggest. As is evident from the plot, the time-averaged mean of this oscillatory variation is zero, but the {\it variance} of ${\cal S}$ has a non-zero finite value that stays nearly constant in the late time limit. One can possibly interpret this as an emergence of correlations in comparison with the state at early times.

One can also directly visualize the evolution of $z$ in the complex plane. The plot for the complex trajectory of $z$ is shown in figure~\ref{o1z}. $z$ starts from zero and after meandering around for a while in the intermediate phase, ultimately ends up, for large $\tau$, on the trajectory $|z|=$ constant, corresponding to a constant mean particle number.  The oscillation of the imaginary part of $z$, as it circles around, is reflected in the late-time variation of the classicality parameter; this can be understood from the relation between $z$ and ${\cal S}$, \eq{J}.

\begin{figure}[h]
\subfigure[ ]{\includegraphics[width=6cm,angle=0.0]{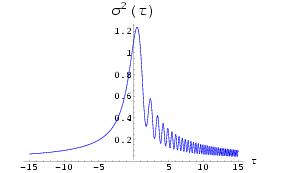}}
\subfigure[ ]{\includegraphics[width=6cm,angle=0.0]{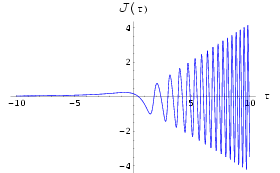}}
\caption{Variation of the functions $\sigma ^2$ and ${\cal J}$ for $\omega_{1}(t)$. The Wigner function starts uncorrelated and peaked on the $p-$axis, and again at late times ends up peaking on the $p-$axis. }
\label{o1W}
\end{figure}

\begin{figure}[h]
\begin{center}
\subfigure[ ]{\includegraphics[width=5cm,angle=0.0]{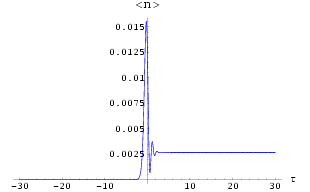}}
\subfigure[ ]{\includegraphics[width=5cm,angle=0.0]{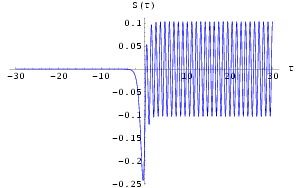}}
\subfigure[ ]{\includegraphics[width=4cm,angle=0.0]{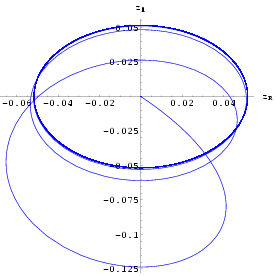}}
\subfigure[ ]{\includegraphics[width=5cm,angle=0.0]{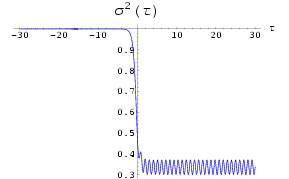}}
\subfigure[ ]{\includegraphics[width=5cm,angle=0.0]{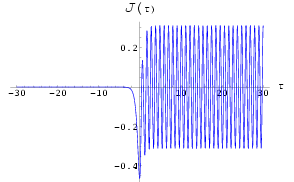}}
\end{center}
\caption{Evolution of the particle number $\langle n \rangle$, classicality parameter ${\cal S}$, complex trajectory of $z$ and the Wigner function variables $\sigma ^2$ and ${\cal J}$ with time for the frequency function $\omega_{2}(t) = \lambda [a + b~ \tanh (\lambda t)]$  ($\epsilon \to 0$ at late times) for the choice $a=2$ and $b=1$. The variation of $\langle n \rangle$, ${\cal S}$ and $z$ are qualitatively similar to those for the frequency function $\omega_1(t)$ considered earlier.}
\label{o2}
\end{figure}

Let us look at the spreading of the Wigner function in the asymptotic limits. As is clear from the plots for $\sigma^2$ and ${\cal J}$ in figure~\ref{o1W}, the Wigner function starts uncorrelated and sharply peaked on the vertical $p$ axis, and again at late times as well, ends up getting concentrated on the $p$ axis. This may be compared with the corresponding classical trajectory, which, in both the asymptotic regions, has the approximate form 
\beq
q_c ~\sim~ \frac{1}{(\lambda + \tau^{2})^{1/4}} \textrm{Re} \l[ {\cal C} e^{i \int^{\tau} \sqrt{\lambda+\tau'^{2}} d \tau'} \r]~\approx~\frac{1}{\sqrt{|\tau|}}\l(1 - \frac{\lambda}{4 \tau^2}\r) \textrm{Re} \l[ {\cal C} e^{i \l(\frac{\tau^{2}}{2} + \frac{\lambda}{2} \ln|\tau| \r)} \r]
\eeq
with the momentum being
\beq
p_c = \dot{q}_c \approx - \frac{q_c}{2 |\tau|}  - \l( \sqrt{|\tau|} + \frac{1}{4 |\tau|^{3/2}} \r) Re \l( i~ {\cal C} e^{i \int^{\tau} \sqrt{1+\tau'^{2}} d \tau'} \r).
\eeq

It is clear that in both limits, $q \to 0$ and the trajectory ends up on the $p$ axis, coinciding with the Wigner function's behavior. The tracking of the classical trajectory by the Wigner function, thus, {\it is not} quite the same as the evolution of the classicality parameter, which, in contrast, displays an asymmetry in time and more closely tracks the particle number. This example also demonstrates that interpreting classicality merely in terms of peaking on the classical trajectory may be suspect, since this can happen even when the oscillator is in a near-vacuum state [which is the case at early times]. But the variation of ${\cal S}$ suggests that the state gets appreciably more correlated in the far future [though remaining oscillatorily zero] {\it in relation} to early times [when it stays very nearly zero]. 
\newline

For the sake of verification of the numerical work, we have repeated the above analysis for another frequency,  given by $\omega_{2}(t) = \lambda [a + b~ \tanh (\lambda t)]$ ($a > b > 0$) which has similar early/late time limits. The analysis of this model proceeds exactly as in the above example. The time evolution is almost identical too, particularly with regard to the particle number and the $q$-$p$ correlation, as is clear from the plots in figure~\ref{o2}. In this case, too, the mean particle number exhibits oscillations that get suppressed as the evolution turns adiabatic for large $\tau$. This shows that our conclusions are fairly generic.
\newline

We now move on to consider a third toy example that is expected to illustrate the {\it alternative} scenario, that of strongly non-adiabatic evolution in the late time limit.
 
\subsection*{Example 2: Non-adiabatic evolution at late times}

Consider the time dependent frequency having the form $\omega_{3}(t) = \lambda / (1+\lambda^{2} t^{2})$. This frequency function presents a sharp contrast to our previous examples. For this case, the adiabaticity parameter $\epsilon(t) = -2 \lambda t$ is small for $|t| \leq \lambda^{-1}$, allowing a reasonable definition of an adiabatic vacuum in the region around $t = 0$. But the magnitude of $\epsilon$ continues to grow linearly with time, and for large $t$, an adiabatic vacuum clearly cannot be defined. Thus, although the usual definition of particle based on an asymptotically adiabatic out- vacuum is not possible here, our choice of the particle number $\langle n \rangle$ provides a very reasonable alternative to quantify the state's particle content in such a situation.  

\begin{figure}[h]
\subfigure[ ]{\label{o3n}\includegraphics[width=6cm,angle=0.0]{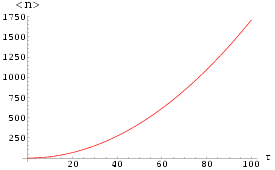}}
\subfigure[ ]{\label{o3S}\includegraphics[width=6cm,angle=0.0]{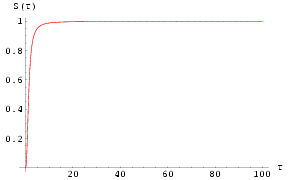}}
\subfigure[ ]{\label{o3z}\includegraphics[scale=0.6]{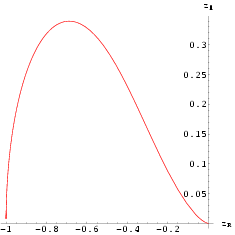}}
\caption{Variation of $\langle n \rangle$, the classicality parameter ${\cal S}$ and $z$ with time for $\omega_{3}(t)$ ($\epsilon \to \infty$ as $t \to \infty$). $\langle n \rangle$ monotonically grows with time and is unbounded; ${\cal S}$ also grows sharply with increasing particle number and saturates at unity at late times. The trajectory of $z$ is confined to a quadrant, and the state ends up at the point $z=-1$ in the $t \to \infty$ limit. }
\label{o3}
\end{figure}

We set the oscillator in the instantaneous ground state at $t=0$ and track its evolution for $t>0$. The plots corresponding to this solution for $z$ are depicted in figures~\ref{o3n}--\ref{o3z}. Our approximations given by Eqs.~(\ref{approx_n}) and (\ref{approx_S}) are borne out, at least qualitatively, by the numerical plots, which show that the particle number $\langle n \rangle$ as well as the phase space correlation $\langle qp \rangle_{{\cal W}}$ grow rapidly without bound at late times. The state can, thus, be regarded as being driven to become strongly correlated with particle creation. This is also in contrast to the [asymptotically] adiabatic evolution for the frequency function $\omega_{1}(t)$ [or of $\omega_{2}(t)$] which was taken up earlier, where ${\cal S}$ stays oscillatorily zero in the absence of particle creation at late times. 

The variation of $z$ in the complex plane, shown in figure~\ref{o3z}, brings out the contrast with the previous case even more vividly. Now $z$ starts from the origin, stays within a quadrant, and ends up going to the limiting value of $z=-1$. The imaginary part of $z$ does not show oscillations, but rises to a maximum before falling off in the large $t$ limit.   

\begin{figure}
\subfigure[ ]{\includegraphics[width=6cm,angle=0.0]{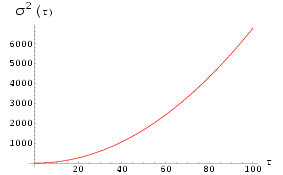}}
\subfigure[ ]{\includegraphics[width=6cm,angle=0.0]{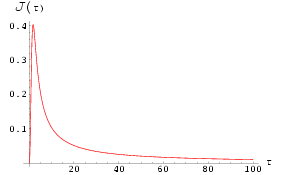}}
\caption{Variation of the functions $\sigma^{2}$ and ${\cal J}$ for $\omega_{3}(t)$. The Wigner function starts uncorrelated in $q$ and $p$ but spread out, and ends up peaking on the $q-$axis at late times. }
\label{o3W}
\end{figure}

Let us also look at the spreading of the Wigner function at late times. From the plots for $\sigma^{2}$ and ${\cal J}$ in figure~\ref{o3W}, it is clear that the Wigner function, starting from being uncorrelated but fairly spread out in phase space at $t=0$, ends up sharply peaking on the $q$ axis as $t \to \infty$. The classical trajectory, in comparison, is given in terms of the exact analytical solution to \eq{mueq}, which is expressible in terms of standard functions~\cite{grad}: 
\beq
q_c = Re[ {\cal A}~ e^{i \sqrt{2}~\tan^{-1} \tau} \sqrt{1 + \tau^{2}}]
\eeq
with 
\beq
p_c = \dot{q}_c = Re \l[ {\cal A}~ \frac{(i \sqrt{2} + \tau)}{\sqrt{1+\tau^{2}}} e^{i \sqrt{2}~\tan^{-1} \tau} \r]
\eeq
so that at late times,
\beq
\frac{p_c}{q_c} = \frac{2 \tau}{1+\tau^{2}} - \frac{2 \sqrt{2}}{1+\tau^{2}} \tan \l( \phi  +  \sqrt{2}~\tan^{-1} \tau \r ) \stackrel{\tau \to \infty} {\longrightarrow} 0. 
\eeq
Thus, the classical trajectory too ends up on the $q$ axis, being tracked by the behavior of the Wigner function. In this instance, the interpretations of the state approaching classicality based on {\it both} peaking on the classical trajectory {\it and} the late time growth of $\langle q p \rangle_{{\cal W}}$ clearly match. 

\begin{figure}
\subfigure[ ]{\label{o3_no}\includegraphics[width=6cm,angle=0.0]{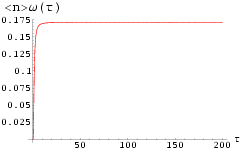}}
\subfigure[ ]{\label{o3_s/t}\includegraphics[width=6cm,angle=0.0]{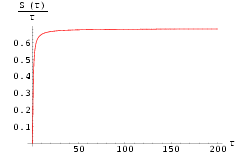}}
\caption{Variation of the functions $\langle n \rangle \omega $ and ${\cal S} / \tau$ for the non-adiabatic case. These quantities saturate at constant values in the $t \to \infty$ limit, in agreement with our late time analytic approximations in \eq{approx_n} and \eq{approx_S}.}
\label{o3_no_s/t}
\end{figure}

The plots in fig.~\ref{o3} follow our expectations based on the approximate analysis carried out in section~\ref{sec:nonadiab} at least at the qualitative level, but we would also like to check if our late time analytic approximations given by Eqs.(\ref{approx_n}) and (\ref{approx_S}) hold. From the expressions for $\langle n \rangle$ and ${\cal S}$, it can be seen that at late times ($t \to \infty$), the variables $\langle n \rangle \omega $ and ${\cal J} \sigma^{2}/ t$ approach constant limiting values. We have plotted the numerical solutions for these quantities as functions of time ($\tau$) in figures~\ref{o3_no} and \ref{o3_s/t}. Sure enough, they reach steady-state values for large $\tau$, in clear agreement with our asymptotic analysis.
 

\section{Discussion}    \label{sec:diss}

In this paper, we have outlined a general formalism to analyze the dynamics of a time dependent oscillator, which provides a reasonable means of quantifying the physical content of the evolving quantum state, and addressing the issue of quantum-to-classical transitions. Among other things, our definition of the time-dependent particle number $\langle n \rangle$ based on instantaneous eigenstates is a very reasonable one to adopt in the absence of adiabatically definable in and out vacua. (The instantaneous particle concept has been considered in field theory in the literature before~\cite{txts1}, and some `difficulties' (the prediction of vastly more particle creation than expected on physical grounds in certain situations etc.) have been pointed out; nevertheless, we believe that in the present context of an oscillator with \emph{general} time dependence, its suitability outweighs its disadvantages.) An interesting feature suggested by our approximate analysis, in particular of adiabatic evolution, is the possible incorporation of information about particle production in the \emph{real} part of the effective Lagrangian, having implications for the issue of studying back-reaction in semiclassical theory, which normally involves considering only the real part of $L_{eff}$. 

We next took up a detailed study of various illustrative toy examples, in an attempt to understand more clearly the connection between particle creation and the approach to classicality based on the Wigner function. We take stock by summarizing the main results of our analysis (in particular the late time behavior of various quantities constructed out of the quantum state, starting from an instantaneous vacuum at the initial moment) of the different examples in tabular form below: 
\vspace{0.5cm}  
\begin{center}
  {\scriptsize
\begin{tabular}{|c|c|c|c|c|c|}
  \hline
& & & & & \\
  & Constant adiabatic &  Constant adiabatic  &  Adiabatic at late times & Adiabatic throughout&  Non-adiabatic at late\\
 & parameter $|\epsilon| \ll 1$ & parameter $|\epsilon| \gg 1$ & ($\epsilon(t) \to 0$ as $t \to \infty$) & (starting from vacuum) & times ($\epsilon(t) \to \infty$ as $t \to \infty$) \\
& & & & &\\
\hline
& & & & &\\
$\omega(t)$ & $\frac{\lambda}{1+\lambda \nu t}$ \quad ($\epsilon=-\nu$) & $\frac{\lambda}{1+\lambda \nu t}$ \quad ($\epsilon=-\nu$) & $\lambda \sqrt{1+\lambda^2 t^2}$  & $|\epsilon(t)|\ll1$ at all times & $\frac{\lambda}{(1+\lambda^2 t^2)}$\\
& & & & & \\
\hline
& & & & & \\
 $ \langle n \rangle $ & Oscillatory at all times & Monotonically & Saturates at a finite & Remains finite and of  & Monotonically increases\\
  & with constant amplitude &  increases with time  & value asymptotically  & order $\epsilon^2$ at all times  & with time \\
& & & & & \\
\hline
& & & & &  \\
 $z(t)$ & A circular trajectory in& Stays within a quadrant & Ends up on a circular & Remains confined to a & Stays within a quadrant\\
  &the complex plane with & and ends up at a point &  trajectory centered on & finite region around the & and ends up at the \\
& $Im z \geq 0$ at all times& with $|z|=1$ &the origin &  origin for general $\epsilon(t)$ & point $z=-1$ \\
& & & & & \\
\hline
& & & & &  \\
${\cal S}(t)$ & Oscillatory at all times & Grows and saturates & Oscillatorily zero with & Remains finite and of & Grows and saturates\\
 & with nearly constant& at unity at late times & finite amplitude at & order $\epsilon$ at all times & at unity at late times\\
& finite amplitude& &  &  &\\
& & & & & \\
\hline
&  &  &  &  & \\
${\cal W} (q,p)$ & Starts spread out & Starts spread out and & Starts peaked on the $p$- & Ends up peaking on either &  Starts spread out and \\
 &  and ends up peaking  & ends up peaking on & axis and again ends up & $q$- or $p$-axis depending on & ends up peaking on the \\
  &on the $q$-axis &the $q$-axis  & peaking on the $p-$axis & late time limit of $\omega(t)$ & $q$-axis  \\
& & & & &\\
\hline

\end{tabular}
}
\end{center}
\vspace{0.5cm}

We began by examining an oscillator with constant adiabaticity parameter, a model that is capable of providing a prototype for adiabatic as well as non-adiabatic evolution. Going from small to large values of $|\epsilon|$, there is a clear variation in the features exhibited by the evolving quantum state. In the adiabatic regime, particle creation is suppressed and the mean particle number, starting from an instantaneous vacuum, maintains an oscillatory profile. The classicality parameter, taken as a measure of phase space correlations, also starts from zero (corresponding to the instantaneous vacuum) and remains bounded and oscillatory. This behavior is sharply contrasted by that in the extreme non-adiabatic case, where the mean number of particles as well as the $q$-$p$ correlation diverge at late times (although the energy saturates at a finite limiting value). We subsequently analyzed three toy models that provide examples of frequency functions which vary adiabatically ($|\epsilon| \to 0$) or non-adiabatically ($|\epsilon| \gg 1$) at late times. These models provide adequate confirmation of the validity and generality of the features revealed by the constant $\epsilon$ case. Our examples also demonstrate that peaking of the Wigner distribution on the classical trajectory is a rather general feature and occurs whenever the frequency goes to zero or infinity, and is quite independent of the particle production (this is particularly clear in the cases of constant $|\epsilon| \ll 1$ and the frequency function $\omega_3(t)$).   

The question of interpreting classical behavior based on the Wigner function involves considering two possible requirements: peaking on the corresponding classical phase space trajectory and emergence of correlation between $q$ and $p$. To repeat once again what we have stated before, our toy examples provide indications that an interpretation based solely on the former requirement can leads to ambiguities. (Another example of such behavior is found in the cosmological context to be discussed in Paper II~\cite{gm}: when evolution of scalar field modes in a de Sitter universe is considered, the Wigner function can be shown to get peaked on the classical trajectory not just at late times, but also in the asymptotic past when the modes are started off in the Bunch-Davies vacuum state.) In light of this fact, we believe that the correlation function ${\cal S}$ comes in as a useful additional variable to quantify classicality. The Wigner function for a vacuum state is expressible as an uncorrelated product of the form $f(q)g(p)$, and for this state, ${\cal S}$ is indeed zero.

Once one chooses to concentrate on the variation of the $q$-$p$ correlation, two possible asymptotic limits can be identified. When the system evolves adiabatically, the correlation function maintains an oscillatory profile. So suppression of particle creation keeps the correlation from varying monotonically (and in fact it can even average out to zero in some cases, like in the example exhibiting late time adiabaticity). If, on the other hand, the system violates adiabaticity at late times (with $\omega,\omega t \to 0$ as $t \to \infty$), large amount of particle production occurs, and this is accompanied by sharp growth in the classicality parameter. This is in reasonable agreement with, and to some extent a generalization of, previous analysis in the literature~\cite{infl2,squeeze,class} associating the notion of classicality with particle creation in the context of inflationary cosmology. 

We would here like to clarify our point of view regarding the interpretation of classicality in relation to the other approaches found in the literature that try to explain emergence of classical properties in terms of the behavior of Wigner functions. It goes without saying that the real world is fundamentally quantum mechanical, and the idea of classical behavior is essentially a matter of interpretation; as a consequence, there is no reason to expect a single criterion to determine classicality to be applicable in all situations. Different criteria are expected to be appropriate depending upon the context. For a normal oscillator, a WKB state is an energy eigenstate, being a solution of the time-independent Schrodinger equation. For the WKB state, it can be shown that the corresponding Wigner function is peaked on the curve $P=f(Q)$, which represents the classical phase space trajectory~\cite{wig2}, and thus shows a precise correlation between position and momentum. The Wigner function in this case does not involve explicit time dependence, and the above-mentioned feature is also naturally independent of time. The other state which has been analyzed in this context is a coherent state which is explicitly time-dependent and for which $\vert \psi \vert ^2$ is actually peaked about a classical trajectory. For such a state, the Wigner function is expressible as an uncorrelated product of gaussians of the form ${\cal W}(q,p,t)=A(q,t)B(p,t)$ and is time-dependent, but the \emph{peak} follows a classical trajectory in phase space~\cite{infl2,wig2}. So, although at any given moment, the Wigner function is not peaked on a classical trajectory (as happens in the case of the WKB stationary state mentioned above), the peak traces out a classical trajectory over time.   

The gaussian state we are working with is, firstly, explicitly time dependent (and so is the corresponding Wigner function) and the `peak', corresponding to the maximum value of ${\cal W}$, remains fixed at the origin ($q=p=0$), since we have chosen a quantum state having zero mean. It is therefore more appropriate to consider the way the Wigner function is \emph{concentrated} in some region of phase space (by, say, tracking the behavior of a Wigner function ellipse, i.e. a contour corresponding to a chosen value of ${\cal W}$). The Wigner distribution is \emph{not} concentrated on the classical trajectory at all times in the manner a stationary WKB state is, and does not track the classical path (unlike the coherent state). But in some particular limit, like at very late times, one \emph{can} make a clear-cut correspondence between the concentration of the Wigner function, based on the orientation and shape of an ellipse, and the behavior of a general classical phase space trajectory in that limit. (Based on a careful analysis of the graphs for the evolution of the Wigner function ellipses and the classical trajectory, such a correspondence does not appear to hold at intermediate times.) It is only in this sense that the Wigner function gets `peaked' on the classical trajectory in some limit, and our interpretations in the various examples studied have been based on making this kind of correspondence.

The essential features revealed by our analysis of the toy models are expected to hold more generally, at least at the qualitative level. In Paper II~\cite{gm}, we shall consider examples from field theory in cosmological and electric field backgrounds.

\section*{ACKNOWLEDGEMENTS}

G.M. is supported by the Council of Scientific $\&$ Industrial Research, India.


\section*{APPENDIX}

\subsection{Derivation of the generating function for $\psi (q,t)$}  \label{app:gfn}

The amplitude for the oscillator to be in the $nth$ instantaneous eigenstate $\phi_{n}(t)$ at time $t$ is non-zero only for even $n$, since $\psi$ is an even function, and is given by
\beq
C_{n} = \int_{-\infty}^{\infty} {dq} \phi_{n}^{*}(q,t) \psi(q,t) = N \l(\frac{m \omega}{\pi}\r)^{1/4}\frac{1}{\sqrt{2^{n} n!}}\int_{-\infty}^{\infty} {dq} H_{n}(\sqrt{m \omega}q) e^{-\l(R+\frac{m \omega}{2}\r)q^{2} + i\int_{t_0}^{t}\l(n + \frac{1}{2} \r) \omega(t)dt }.
\eeq
This can be evaluated by making use of the following generating function for Hermite polynomials~\cite{arf}:
\beq
e^{-t^{2} + 2tx} = \sum_{n=0}^{\infty} \frac{H_{n}(x) t^{n}}{n!}.
\eeq
Substituting $x=\sqrt{m \omega}q$, multiplying both sides by $\exp[-(R+ m \omega/2)q^{2}]$, and then integrating w.r.t. $q$ finally gives
\beq
\sqrt{\frac{\pi}{(R+\frac{m \omega}{2})}} e^{t^2\l(\frac{m \omega}{R+m \omega/2} - 1\r)} = \sum_{n=0}^{\infty} \frac{t^n I_{n}}{n!}.
\eeq
Equating coefficients of equal powers of $t$ on both sides gives
\beq
I_{2n} = (2n)!\sqrt{\frac{\pi}{(R+\frac{m \omega}{2})}} \frac{\l( \frac{m \omega}{R+m \omega/2} - 1 \r)^n}{n!} \qquad (n = 0, 1, 2,...)
\eeq
and
\beq
C_{2n} =  N \l(\frac{m \omega}{\pi}\r)^{1/4}\frac{I_{2n}}{\sqrt{2^{2n} (2n)!}}~e^{i\int_{t_0}^{t}\l(2n + \frac{1}{2} \r) \omega(t)dt}    \label{C_2n_Spic}
\eeq
while $I_{n}$ vanishes for odd $n$. Therefore, the probability for the oscillator to be in the eigenstate $\phi_{2n}$ at time $t$ is simply
\beq
P_{2n} = |C_{2n}|^{2}
= P \frac{(2n)!}{(n!)^2} \frac{|z|^{2n}}{2^{2n}},   
\eeq
where
\beq
z =  \l( \frac{\omega + \frac{i}{\mu}\frac{d \mu }{dt}}{\omega -\frac{i}{\mu}\frac{d\mu}{dt}}\r)
\eeq
using \eq{z}, and
\beq
P = \frac{|N|^2 \sqrt{\pi m \omega}}{|R+\frac{m \omega}{2}|}.
\eeq
The generating function for this probability distribution is defined as 
\beq
G(x) \equiv \sum_{n=0}^{\infty} P_{2n} x^{n} =  P \sum_{n=0}^{\infty} \frac{(2n)!}{(n!)^2} \frac{|z|^{2n}}{2^{2n}} x^{n}.
\eeq
It can be evaluated using the following relation~\cite{arf}:
\beq
 \sum_{n=0}^{\infty} \frac{(2n)!}{(n!)^2}\frac{X^n}{2^{2n}} = \l( 1 - X \r)^{-1/2}
\eeq
which gives (setting $x \vert z \vert ^2 = X$):
\beq
G(x) =  P \l( 1-x|z|^2 \r)^{-1/2}.  
\eeq
And finally, since the total probability given by $G(1)$ is unity, we have
\beq
P = \sqrt{1-|z|^2}.    \label{prob1}
\eeq

\subsection{Some properties of the probability distribution for the gaussian state}  \label{app:p_d}

Consider the probability distribution of the particle number for our gaussian quantum state, given by \eq{genfn1}. Setting $|z|^2 = \exp{(-b)}$ and $x=\exp{(-\lambda)}$, the generating function in \eq{genfn1} can be written as
\beq
G(\lambda) = \sqrt{\frac{1-e^{-b}}{1-e^{-b - \lambda}}} .    \label{genfn2}
\eeq
(It may be remembered that this is a generating function for $\it pairs$ of particles, i.e. the $k$th term in the series expansion in powers of $\exp{(-\lambda)}$ gives the probability for $k$ $\it pairs$.) In terms of $b$, the mean particle number has the form $\langle n \rangle = (\exp{(b)}-1)^{-1}$, which is same as for a Planckian distribution specified by a temperature $b^{-1}$. It's also straightforward to show that
\beq
\langle n^2 \rangle =  4 \sum_{k=0}^{\infty} k^2 P_{2k} = 4 \frac{d^2 G(\lambda)}{d \lambda^2}_{|\lambda=0} = 3 \langle n \rangle ^2 + 2 \langle n \rangle
\eeq
and so the dispersion in the particle number is given by
\beq
(\Delta n )^2 = \langle n^2 \rangle - \langle n \rangle^{2} = 2 \langle n \rangle \l( \langle n \rangle + 1 \r).
\eeq
Let us also look at a thermal state in comparison, which is given by the following distribution:
\beq
\bar{P}(n) = \mathcal{N} e^{- b n} \qquad (n = 0, 1, 2,...)
\eeq
with the normalization $\mathcal{N}=(1-e^{-b})$ fixed by requiring the total probability to be unity. This distribution is described by the generating function
\beq
\bar{G}(\lambda) = \sum_{n=0}^{\infty} \bar{P}(n) e^{-\lambda n} = \l( \frac{1-e^{-b}}{1-e^{-b - \lambda}} \r). \label{genfn3}
\eeq
from which it is clear that the generating functions of the two distributions are related in a simple manner as follows:
\beq
G(\lambda) = \sqrt{\bar{G}(\lambda)}.
\eeq
For the thermal distribution, the first two moments are given by
\beq
\langle n \rangle = \frac{1}{e^{b}-1} \quad,\quad \langle n^2 \rangle =  \frac{1}{e^{b}-1} +  \frac{2}{\l( e^{b}-1 \r) ^2}
\eeq
and hence
\beq
(\Delta n )^2 = \langle n \rangle \l( \langle n \rangle + 1 \r)
\eeq
which is the standard result~\cite{squeeze}. The distribution in \eq{genfn2}, thus, has a dispersion that differs by only a factor of 2 from that for the thermal state. This fact has been pointed out in the literature before (see for e.g.~\cite{grishchuk}), but its significance, if any, is unclear.

\subsection{Connecting with the Heisenberg picture}  \label{app:hpic}

For completeness and for reference, we give below an outline of the analysis of the time-dependent oscillator in the Heisenberg picture.

The starting point is the Hamiltonian, defined as
\beq
H(q,p)~=~\frac{1}{2} \l(p^2 + \omega^2(t) q^2 \r) \quad \textrm{with} \quad p = \dot{q}
\eeq
and $q$ satisfies the equation of motion
\beq
\ddot{q}(t) + \omega^{2}(t) q(t) = 0.
\eeq
$q$ is elevated to the status of an operator, and we write it as
\beq
q(t) = a f(t) + a^{\dagger} f^* (t)   \label{q_h_pic}
\eeq
where $a,a^{\dagger}$ are the time independent creation and annihilation operators, and $f(t)$ satisfies the same equation as $q(t)$. Further, requiring $[q,p]=i$ and the ladder operators to satisfy the commutation relation $[a,a^{\dagger}]=1$, one can show that
\beq
{\dot f}^{*}f - {\dot f} f^{*} = i.
\eeq
Next, the function $f$ is written in terms of two new functions $A(t)$ and $B(t)$ in the following form:
\beq
f(t) = A(t) e^{-i\rho(t)} + B(t)  e^{i\rho(t)} \qquad (\dot{\rho} = \omega)  \label{f_h_pic}
\eeq
Since two functions have been introduced, an additional constraint needs to be specified to fix the form of these functions. We impose the requirement that $\dot{f}(t)$ be of the form 
\beq
\dot{f}(t) = -i \omega A(t) e^{-i\rho(t)} + i \omega B(t)  e^{i\rho(t)}   \label{dot_f_h_pic}
\eeq
i.e. the derivative of $f$ have the same form as it would if $A,B$ were to be independent of time. This gives the following constraint on $A,B$:
\beq
\dot{A} e^{-i \rho} + \dot{B} e^{i \rho} = 0.    \label{eq:1}
\eeq
Further, substituting the expression in \eq{f_h_pic} in the equation for $f$ and using (\ref{dot_f_h_pic}) gives
\beq
\dot{A} e^{-i \rho} - \dot{B} e^{i \rho} = \frac{\dot \omega}{\omega} \l( B e^{i \rho} - A e^{-i \rho} \r)   \label{eq:2}
\eeq
Using \eq{eq:1} and \eq{eq:2}, one ends up with the following pair of coupled equations for $A$ and $B$: 
\br
\dot{B} + \frac{\dot{\omega}}{2 \omega}B &=& \frac{\dot{\omega}}{2 \omega} A e^{-2 i \rho},  \nonumber \\
\dot{A} + \frac{\dot{\omega}}{2 \omega}A &=& \frac{\dot{\omega}}{2 \omega} B e^{2 i \rho}.  \label{eq:3}
\er
Let $A(t)= \alpha(t)/\sqrt{2 \omega}$ and $B(t)= \beta(t)/\sqrt{2 \omega}$. Then \eq{eq:3} reduce to
\br
\dot{\alpha} &=& \frac{\dot{\omega}}{2 \omega} \beta e^{2 i \rho}~~\text{and} \nonumber \\
\dot{\beta} &=& \frac{\dot{\omega}}{2 \omega} \alpha e^{-2 i \rho}  \label{h_pic_eqs}.
\er
Given the functional form of $\omega(t)$, these coupled equations specify the complete time evolution of the system starting from any initial condition on $\alpha$ and $\beta$.

\eq{q_h_pic} is next rewritten in terms of $\alpha,\beta$: 
\br
q(t) = a f(t) + a^{\dagger} f^{*}(t) &=& \l( \frac{\alpha(t)}{\sqrt{2 \omega}} a + \frac{\beta^*(t)}{\sqrt{2 \omega}} a^{\dagger}  \r) e^{-i\rho} + \l( \frac{\beta(t)}{\sqrt{2 \omega}} a + \frac{\alpha^*(t)}{\sqrt{2 \omega}} a^{\dagger}  \r) e^{i\rho} \nonumber \\
&\equiv& \frac{a(t) e^{-i\rho} + a^{+}(t) e^{i\rho}}{\sqrt{2 \omega}}.
\er
The last line above defines for us the time-dependent creation and annihilation operators as
\beq
a(t) = \alpha(t) a + \beta^{*}(t) a^{\dagger}~~,~~a^{\dagger}(t) = \alpha^{*}(t) a^{\dagger} + \beta(t) a  \label{def_a(t)}
\eeq
and these imply
\beq
a = \alpha^{*}(t) a(t) - \beta^{*}(t) a^{\dagger}(t)~~,~~a^{\dagger} = \alpha(t) a^{\dagger}(t) - \beta(t) a(t). \label{def_a}
\eeq
($a(t),a^{\dagger}(t)$ are defined in such a way that for the constant frequency case, they remain independent of time.) 

Let $|n,t \rangle$ stand for the n-particle state at time $t$. The vacuum at time $t$ is defined by the condition $a(t) |0,t \rangle = 0$. Let $\alpha(0)=1$ and $\beta(0)=0$ so that $a$ and $a^{\dagger}$ coincide with the ladder operators at $t=0$. One can then expand the vacuum state at $t=0$ (the {\it in} vacuum) in terms of the basis of states defined at $t$:
\beq
|0,0 \rangle = \sum_{n=0}^{\infty} C_n |n,t \rangle.
\eeq 
Operating with $a$ on both sides and using the relation in \eq{def_a} gives
\beq
C_1 = 0~;~C_{2m+1} = 0 \qquad (m = 0, 1, 2,...)
\eeq
and
\beq
C_{2m} = \frac{\beta^*}{\alpha^*}\sqrt{\frac{2m-1}{2m}}C_{2m-2} \qquad (m = 1, 2,...).
\eeq
The recurrence relation above allows one to express $C_{2m}$ in terms of $C_0$, and thus
\beq
|0,0 \rangle = C_{0} \sum_{m=0}^{\infty} \l(\frac{\beta^*}{\alpha^*}\r)^{m} \frac{\sqrt{(2m)!}}{2^{m}m!} |2m,t \rangle.
\eeq 
Using the relation $\l(a^{\dagger}(t)\r)^{2m}|0,t\rangle = \sqrt{(2m)!}|2m,t\rangle$, the expression above can be rewritten in the following form:
\beq
|0,0 \rangle = C_{0} \sum_{m=0}^{\infty} \frac{1}{m!} \l(\frac{\beta^*}{2\alpha^*}a^{\dagger 2}(t)\r)^{m}|0 , t \rangle \equiv C_0 \exp \l( \frac{\beta^*}{2\alpha^*}a^{\dagger 2}(t) \r) |0,t \rangle.    \label{C_2m_Hpic}
\eeq
The magnitude of $C_0$ can be set by normalization:
\beq
\langle 0,0 | 0,0 \rangle \equiv \sum_{m=0}^{\infty} |C_{2m}|^{2} = 1 ~\Longrightarrow~ \frac{|C_0|^{2}}{\sqrt{1-\l|\frac{\beta}{\alpha}\r|^2}} = 1 ~\Longrightarrow~ |C_0|^{2}|\alpha| = 1
\eeq
so $|C_0|=|\alpha|^{-1/2}$. 

One can compare the coefficient in \eq{C_2m_Hpic} with the expression for $C_{2m}$ derived earlier in the Schrodinger picture, \eq{C_2n_Spic}, to make the correspondence $z \exp(2 i \rho) = \beta^{*}/\alpha^{*}$ with $\dot{\rho}=\omega$. This relation can also be deduced by working out the equation satisfied by $(\beta^* / \alpha^*)\exp(-2 i \rho)$ using \eq{h_pic_eqs}, which turns out to be the same as the $z$ equation in (\ref{eq:z}).

\subsection{Late time analytic approximation for $z$ when $\epsilon(t) \to \infty$}  \label{app:a_a}

Making the substitution $z=(1+\Delta)\exp(i\theta)$~(with $\Delta$ real) in the equation for $z$, eq.(\ref{eq:z}), the following coupled equations for $\Delta$ $\&$ $\theta$ are obtained:
\br
\omega \theta ' + \l[ 1 + \frac{\Delta^2}{2 (\Delta + 1)} \r] \sin \theta + \frac{2}{\epsilon} = 0, \nonumber \\
\omega \Delta ' + \l( \Delta + \frac{\Delta^{2}}{2} \r) \cos \theta = 0   \label{eq:d_t_1}
\er
where the prime ($'$) denotes differentiation w.r.t. $\omega$. We are seeking an approximate late time solution with $\Delta \to 0$ as $\omega \to 0$ (i.e. $t \to \infty$). Assuming that in this limit, the O($\Delta^{2}$) terms in eq.(\ref{eq:d_t_1}) can be dropped, we are left with
\br
\omega \theta ' + \sin \theta + \frac{2}{\epsilon} \approx 0, \nonumber \\
\omega \Delta ' +  \Delta \cos \theta \approx 0.   \label{eq:d_t_2}
\er
Let $\sin \theta \to 0$ as $\omega \to 0$, i.e. $\theta \to n \pi$ ($n~\epsilon~{\cal I}$). This would give $\cos \theta \to \pm 1$. Consider the upper limit first. Then $\theta \to n \pi$ with even $n$. This gives
\br
\omega \Delta ' +  \Delta \approx 0  \nonumber \\
\Longrightarrow~~\Delta \sim \frac{c_{1}}{\omega}       
\er
which, in the $\omega \to 0$ limit, blows up, and we run into an inconsistency with the dropping of the O($\Delta^{2}$) terms earlier. Consider therefore the other limit, i.e. $\cos \theta \to -1$. Let $\theta = n \pi - \phi$ ($n$ is odd). Eqs.(\ref{eq:d_t_2}) reduce to
\br
-\omega \phi ' + \phi \approx -\frac{2}{\epsilon}~, \nonumber \\
\omega \Delta '  - \Delta \approx 0.
\er    
The second equation above can be integrated to give $\Delta \sim c_{1} \omega$ which yields the right late time limit. The first equation can also be solved, and gives
\br
\frac{\phi}{\omega} ~\approx~ c_{0} + \int^{\omega} \frac{2}{\epsilon~\omega^{2}} d \omega  ~\equiv~ c_{0} + 2 t 
\er
using the relation $\epsilon = \dot{\omega}/\omega^{2}$.


\end{document}